\documentclass[12pt]{article}
\usepackage{a4wide,graphics,graphicx,amsmath,amssymb,cite,nicefrac}
\usepackage[verbose]{wrapfig}
\usepackage{authblk,hyperref}
\usepackage{url,amsfonts}
\usepackage[makeroom]{cancel}
\catcode`@=11 \@addtoreset{equation}{section} \catcode`@=12

\begin{document}

\begin{titlepage}%1
\begin{center}
%\hfill DFPD/2017/TH/\\

\vskip 1.0cm

{\bf \huge Penrose limits  in non-Abelian T-dual \\ \vskip .5cm of  Klebanov-Tseytlin Background}

\vskip 1.0cm

{\bf \large Sourav Roychowdhury${}^{1}$ and Prasanta K. Tripathy${}^{2}$}

\vskip 30pt
 {\it ${}^1$Chennai Mathematical Institute, \\
SIPCOT IT Park, Siruseri 603 103, India}\\\vskip 5pt
{\it ${}^2$%
Department of Physics, Indian Institute of
Technology
Madras,  \\ Chennai
600 036, India}\\

\vskip 10pt

\texttt{%
souravroy@cmi.ac.in},
\texttt{%
prasanta@iitm.ac.in}

\end{%
center}

\vskip 1.5cm

\begin{%
center} {\bf ABSTRACT}\\[3ex]\end{center}%

In this paper we consider the Klebanov-Tseytlin background and its non-Abelian
T-dual geometry along a suitably chosen $SU(2)$ subgroup of isometries. We 
analyse the Penrose limits along various null geodesics of both the geometries.
We observe that, the Klebanov-Tseytlin geometry does not admit any pp-wave 
solutions. However, the T-dual background gives rise to pp-wave solution upon 
taking the Penrose limit along some appropriate null geodesic. We comment
on the possible gauge theory dual for our pp-wave background.

%\today

%\end{center}%

%\noindent

%
\vfill

%\July 2008

\end{titlepage}

\newpage % \setcounter{page}{1} \numberwithin{equation}{section}
%\pagestyle{plain}
%\tableofcontents

\section{\label{Intro}Introduction}

String theory on pp-wave background is being analysed extensively during 
the past several decades because they are endowed with a number of unique 
features \cite{Kowalski-Glikman:1984qtj,Chrusciel:1984gr,Figueroa-OFarrill:2001hal,Blau:2001ne,blau.m}. These pp-wave solutions appear as Penrose limits of various 
supergravity backgrounds in ten and eleven dimensions \cite{Blau:2002dy,Berenstein:2002jq,Sadri:2003pr}. They provide exact
string theory backgrounds to all orders in $\alpha'$ as well as $g_s$ \cite{Amati:1988sa,Horowitz:1989bv}.
In the context of AdS/CFT correspondence these backgrounds give rise to 
the so called BMN sector with large $R$-charge of the dual $\mathcal{N}=4$ 
superconformal theory in four dimensions \cite{Berenstein:2002jq}. The AdS/CFT correspondence is 
used to construct interacting string states from perturbative gauge 
theory \cite{Berenstein:2002jq}.

Recently pp-wave backgrounds have been constructed from non-Abelian T-dual
geometries of various supergravity theories. Non-Abelian T-duality has turned
out to be a wonderful tool to construct new supergravity backgrounds from
known ones. This is a nontrivial generalization of the conventional T-duality
where a non-Abelian isometry group is used for dualization \cite{delaOssa:1992vci}. However, these
non-Abelian T-dualities are not symmetries of the full string theory \cite{Giveon:1993ai}. They are 
used to relate the low energy supergravity theories among each other. 
Originally non-Abelian T-duality was formulated for the NS sector of supergravity theories.
Subsequently, this formalism has been generalized to include the RR fields \cite{Sfetsos:2010uq}.
This, in turn played a crucial role in relating different supergravity 
backgrounds among each other. Several examples of new supergravity backgrounds
have also been constructed using the non-Abelian T-duality \cite{Itsios:2012zv,Itsios:2012dc,Macpherson:2014eza,Lozano:2011kb,Dimov:2015rie,Barranco:2013fza,Kooner:2014cqa,Dimov:2016rff}. 

Of particular interest in the present context is the impact of these developments
in understanding several aspects of AdS/CFT correspondence \cite{Itsios:2013wd,Araujo:2015npa,Whiting,Lozano:2016kum,Lozano:2016wrs,Itsios:2017nou,Itsios:2017cew,Gaillard:2013vsa}. This has opened up 
the possibility of constructing several new CFT duals corresponding to these 
non-Abelian T-dual geometries. Relationships between a number of these dual 
geometries with the Penrose limits \cite{PenroseLimit} of some of the prevailing supergravity 
backgrounds have also been revealed \cite{Dimov:2016rff,Lozano:2016kum,Lozano:2016wrs}. An important development in this context is
the non-Abelian T-dual of type-$IIB$ supergravity compactified on certain orbifolds
of $AdS_5\times S^5$ \cite{Itsios:2017nou}. It has been shown that this geometry indeed admits plane
wave solutions upon taking the Penrose limit along appropriate null geodesics \cite{Itsios:2017nou}.
A candidate for the field theory dual of this geometry has also been proposed.
These developments have further been generalized for the non-Abelian T-duals 
of the Klebanov-Witten background, which results in placing a stack of $D3$ branes 
near a conifold singularity. The corresponding supergravity background, 
$AdS_5\times T^{1,1}$, is obtained by blowing up the singularities of $AdS_5\times
S^5$ orbifolds \cite{Klebanov:1998hh,Klebanov:1999tb}. An appropriate $SU(2)$ subrgoup of isometries of $T^{1,1}$ can
be used to obtain the non-Abelian T-dual geometry \cite{Itsios:2017cew}. These dual goemetries also 
give rise to pp-wave solutions upon considering the Penrose limits along 
appropriate null geodesics \cite{Roychowdhury:2019kqr}.

Placing a stack of $M$ fractional along with $N$ regular $D3$ branes at the conifold 
singularity gives rise to a $\mathcal{N}=1$ supersymmetric $SU(N+M)\times SU(N)$ gauge 
theory. The gravity dual is a nontrivial modification of the $AdS_5\times T^{1,1}$ 
background resulting the well-known Klebanov-Tseytlin geometry \cite{Klebanov:2000nc}. The goemetry 
admits non-Abelian isometries, an $SU(2)$ subgroup of which is used to 
construct a non-Abelian T-dual background. This gives rise to a new massive type-$IIA$ 
supergravity background \cite{Itsios:2012zv,Itsios:2013wd}. In the present work we will analyse the Penrose 
limits of this massive type-$IIA$ background in addition to the original 
Klebanov-Tseytlin background. We will show that, for the type-$IIA$ theory,
 the resulting
background indeed admits a pp-wave solution. The plan of this paper is as 
follows. In the next section we will review the Klebanov-Tseytlin background
and its non-Abelian T-dual. Subsequently, in \S3 we will analyse the Penrose 
limits and obtain pp-wave solution. In \S4 we consider the supersymmetry
analysis and show that the resulting pp-wave background preserves 16
supercharges. Finally, we will discuss some aspects of the dual 
quiver theory before summarising the results.

\section{Klebanov-Tseytlin Background} \label{original background} 

Placing a stack of $N$ regular and $M$ fractional branes at a conifold singularity
modifies the spacetime geometry in a non-trivial way. The gravity dual of this non-conformal
$SU(N+M)\times SU(N)$ gauge theory has been accomplished in a pioneering work by Klebanov
and Tseytlin \cite{Klebanov:2000nc}. The geometry of the resulting supergravity  
background is given by  
\begin{eqnarray} \label{KT}
ds^2 =  H(r)^{-\frac{1}{2}} \ \eta_{\mu\nu} dx^\mu dx^\nu + H(r)^{\frac{1}{2}} \Big(dr^2 + r^2 ds_{T^{1,1}}^2\Big) \ .
\end{eqnarray}
Here we will use the conventions of \cite{Kruczenski:2003wz}. The warp factor is 
given by 
\begin{eqnarray} \label{h}
H(r) =  \frac{1}{r^4} \Bigg[R^4 + 2L^4 \Bigg(\ln\bigg(\frac{r}{r_0}\bigg) + \frac{1}{4}\Bigg)\Bigg]  \ ,
\end{eqnarray}
with $\eta_{\mu\nu}$ denoting the stranded $(3+1)$-dimensional 
Minkowski metric. The metric of  $T^{1,1}$ is given by \cite{Klebanov:1999tb}
\begin{eqnarray} \label{T 1,1}
ds^2_{T^{1,1}}= \lambda_{1}^2\ d\Omega^2_{2} \big(\theta_{1},\phi_{1}\big) + \lambda_{2}^2\ d\Omega^2_{2} \big(\theta_{2},\phi_{2}\big)
+ \lambda^2\big(d\psi + \cos\theta_{1} d\phi_{1} + \cos\theta_{2} d\phi_{2}\big)^2 \ .
\end{eqnarray}
In the above, $d\Omega_2^2(\theta,\phi)$ denotes the round metric on a two-sphere.
For $T^{1,1}$ the parameters $\lambda,\lambda_1,\lambda_2$ take the numerical values
  $\lambda_{1}^2=\lambda_{2}^2=\frac{1}{6}, \lambda^2=\frac{1}{9}$.

In addition, we need to specify non-vanishing background fields in NS-NS and RR 
sectors. The background NS-NS two form field $B_2$ has the expression
\begin{eqnarray} \label{KT B2}
B_2 %=& \frac{L^2}{3}  \ln\big(\frac{r}{r_0}\big) \Big(\sin\theta_1 d\theta_1 \wedge d\phi_1 - \sin\theta_2 d\theta_2 \wedge d\phi_2\Big) \ , \nonumber\\
= \frac{T(r)}{6\sqrt{2}} \Big(\sin\theta_1 d\theta_1 \wedge d\phi_1 - \sin\theta_2 d\theta_2 \wedge d\phi_2\Big) \ ,
\end{eqnarray}
with the corresponding field strengths
\begin{eqnarray} 
H_3 = \frac{L^2}{3r} \Big(\sin\theta_1 dr \wedge d\theta_1 \wedge d\phi_1 - \sin\theta_2 dr \wedge d\theta_2 \wedge d\phi_2\Big) \ .
\end{eqnarray}

The non-vanishing RR fields strengths $F_3$ and $F_5$ are given respectively by
\begin{eqnarray} 
F_3 = \frac{P}{18\sqrt{2}} \Big(d\psi + \cos\theta_{1} d\phi_{1} + \cos\theta_{2} d\phi_{2}\Big) \wedge \Big(\sin\theta_1 d\theta_1 \wedge d\phi_1 - \sin\theta_2 d\theta_2 \wedge d\phi_2\Big) \ , % \nonumber\\
\end{eqnarray}
and
\begin{eqnarray} 
F_5= \big(1 + *_{10} \big) K(r) {\text{Vol}} \big(T^{1,1}\big) \ .
\end{eqnarray}
Here, $*_{10}$ denotes the Hodge dual with respect to the ten dimensional metric 
\eqref{KT}. For convenience, in the above we have used the notation 
\cite{Klebanov:2000nc}
\begin{eqnarray} \label{KT B21}
&P=& \frac{L^2}{ g_s} \ 2\sqrt{2}  \ ,  \nonumber\\
&T(r)=& 2\sqrt{2} L^2 \ln\bigg(\frac{r}{r_0}\bigg)       \ ,   \nonumber\\
&K(r)=&  \frac{r^4}{30} \ H(r) \Bigg[1 - \frac{L^4}{2r^4 H(r)}\Bigg]   \ .
\end{eqnarray}
The numbers $N$ of regular $D3$ branes, and $M$ of fractional $D3$ branes 
corresponds respectively to the flux of $F_5$ and $F_3$. It is important to note that, the constant $P$ is proportional to the number $M$ of fractional $D3$ branes.

%from above we have $\frac{1}{12} H_3^2 = \frac{4L^4}{r^6} H(r)^{-\frac{3}{2}}= R$. 
%\begin{eqnarray} 
%&H_3 \wedge F_3 =& \frac{2L^4}{27 r g_s} \sin\theta_1 \sin\theta_2 \ d\psi \wedge d\theta_1 \wedge d\phi_1 \wedge d\theta_2 \wedge d\phi_2 \wedge dr \ , \nonumber\\
%&d \Big(\star F_5\Big) =&  5! \ \partial_r K(r) \lambda \lambda_1^4 \sin\theta_1 \sin\theta_2  \ d\psi \wedge d\theta_1 \wedge d\phi_1 \wedge d\theta_2 \wedge d\phi_2 \wedge dr\ .  \nonumber\\
%\end{eqnarray}
%\begin{eqnarray} 
%d \Big(\star F_5\Big) - H_3 \wedge F_3  = 0 \ , \  \partial_r K(r) = \frac{L^4}{15r} \ . 
%\end{eqnarray}

%\subsection{Non-Abelian T-dual of Klebanov-Tseytlin background} \label{natd} 

We will now consider the non-Abelian T-dual of the Klebanov-Tseytlin background.
The non-Abelian T-duality with respect to an $SU(2)$ isometry has been obtained 
in \cite{Itsios:2012zv, Itsios:2013wd}. The corresponding metric of the T-dual 
geometry is given by 
\begin{eqnarray} \label{NATD-dual metric}
&d\hat{s}^2_{\text{NATD}}=&   H(r)^{-\frac{1}{2}}  \ \eta_{\mu\nu} dx^{\mu} dx^{\nu}  + H(r)^{\frac{1}{2}} \Big(dr^2 + \frac{1}{6} r^2  d\Omega_2^2  (\theta_1, \phi_1) \Big) \ \nonumber\\
&& + \ \frac{1}{2r^2 \Delta \ H(r)^{\frac{1}{2}}} \Bigg[12 r^4 H(r) v_2^2 \sigma_{\hat{3}}^2 + 12 \Big(r^4 H(r) + 27 v_2^2\Big) dv_2^2  \nonumber\\
&& \ + \ 9\Big(2r^4 H(r) + \mathcal{V}^2\Big) dv_3^2 + 108 \mathcal{V} v_2 \ dv_2 dv_3 \Bigg] \ ,
\end{eqnarray}
where, the one form $\sigma_{\hat 3}$ is defined as 
\begin{equation}
\sigma_{\hat{3}} = d\psi + \cos\theta_1 d\phi_1 \ ,
\end{equation}
and the functions $\Delta$ and $\mathcal{V}$ are given by 
\begin{eqnarray} \label{delta, gamma}
\Delta = 2r^4 H(r) +  \mathcal{V}^2 + 54v_2^2 \ ,  \ \mathcal{V} = 6v_3 + 2\sqrt{2} L^2 \ln\bigg(\frac{r}{r_0}\bigg)   \ .
\end{eqnarray}
The expression for the background NS-NS two form $\hat{B}_2$ and dilaton $\hat\Phi$ 
of the dual geometry are given by
\begin{eqnarray} \label{NATD dila, b2}
&\hat{B}_2& =   \frac{L^2}{3} \ln\bigg(\frac{r}{r_0}\bigg)   \sin\theta_1 d\theta_1 \wedge d\phi_1 + \frac{3\sqrt{2}}{\Delta} \ \mathcal{V} v_2 \sigma_{\hat{3}} \wedge dv_2  + \frac{1}{\sqrt{2}\Delta} \Big(2r^4 H(r) +  \mathcal{V}^2\Big) \sigma_{\hat{3}} \wedge dv_3 \ ,\nonumber\\
&e^{-2\hat{\Phi}}& = \frac{1}{81 g_s^2}  \ r^2 H(r)^{\frac{1}{2}} \ \Delta \ .
\end{eqnarray}
The field strengths corresponding to the RR sector are
\begin{eqnarray} \label{NATD-dual RR}
&\hat{F}_0=& -  L^2 \ \frac{2\sqrt{2}}{9 g_s} \ , \nonumber\\
&\hat{F}_2=&   \frac{1}{162\sqrt{2} \ g_s} \Bigg[\frac{r^4}{5} \ H(r) \bigg(1 - \frac{L^4}{2r^4 H(r)}\bigg) + L^2 \ 6\sqrt{2}  \bigg(6v_3 + 2\sqrt{2} L^2 \ln\Big(\frac{r}{r_0}\Big)  \bigg)\Bigg] \sin\theta_1 d\theta_1 \wedge d\phi_1   \nonumber\\
&&-  \ L^2  \frac{4}{3 g_s} \ \frac{6v_3 + 2\sqrt{2} L^2 \ln\Big(\frac{r}{r_0}\Big)  }{\Delta} \ v_2 \ \sigma_{\hat{3}} \wedge dv_2 +  L^2  \ \frac{12}{ g_s}  \frac{v_2^2}{\Delta} \ \sigma_{\hat{3}} \wedge dv_3 \ , \nonumber\\
&\hat{F}_4=&   \frac{v_2}{18 \Delta \ g_s} \sin\theta_1 d\theta_1 \wedge d\phi_1 \wedge d\psi \wedge \Bigg[\Bigg(-18 \sqrt{2} L^2 \bigg(6v_3 + 2\sqrt{2} L^2 \ln\Big(\frac{r}{r_0}\Big)   \bigg)  \nonumber\\
&& - \frac{3r^4}{5} \ H(r) \Big(1 - \frac{L^4}{2r^4 H(r)}\Big)\Bigg) v_2 dv_3  +\ 2 \Bigg(-2\sqrt{2} L^2 r^4 H(r) + \frac{r^4}{30} \bigg(6v_3  \nonumber\\
&& + 2\sqrt{2} L^2 \ln\Big(\frac{r}{r_0}\Big) \bigg) H(r) \Big(1 - \frac{L^4}{2r^4 H(r)}\Big) - 54 \sqrt{2} L^2  v_2^2 \Bigg) dv_2 \Bigg] \ . 
\end{eqnarray}

\section{The Penrose Limits}

In this section we will study Penrose limits for both of the above backgrounds.
We will first consider the original type-$IIB$ background. The Penrose limit along 
a suitable null geodesics for this background has already been studied in 
\cite{PandoZayas:2002dso}. Here we will first outline the main result of this work. 
Considering the motion of a massless particle in the $(r,\psi)$ 
plane of the background results the following geometry 
\begin{eqnarray} \label{pp in 2b}
&& ds^2 = 2 dudv  + \frac{r^2}{\sqrt{1 + P \ln\Big(\frac{r}{r_0}\Big)}} \ dx_3^2  +  \sqrt{1 + P \ln\Big(\frac{r}{r_0}\Big)} \ \Bigg[1 - \frac{\mu^2 r^2}{1 + P \ln\Big(\frac{r}{r_0}\Big)}\Bigg] dx^2  \nonumber\\
 &&+  \sqrt{1 + P \ln\Big(\frac{r}{r_0}\Big)} \Bigg[dr_1^2 + r_1^2 d\phi_1^2 + dr_2^2 + r_2^2 d\phi_2^2\Bigg] - \frac{\mu^2}{\sqrt{1 + P \ln\Big(\frac{r}{r_0}\Big)}} \Big(r_1^2 + r_2^2\Big) du^2 . \ {~~} %\nonumber\\
\end{eqnarray}
The background gauge fields behave as 
\begin{eqnarray} \label{2b fields}
&B_2 \sim& P \ln \Big(\frac{r}{r_0}\Big) \Big(dr_1 \wedge r_1 d\phi_1 - dr_2 \wedge r_2 d\phi_2\Big) \ ,  \nonumber\\
&F_3 \sim& P \dot{\psi} \ du \wedge \Big(dr_1 \wedge r_1 d\phi_1 - dr_2 \wedge r_2 d\phi_2\Big) \ ,  \nonumber\\
&F_5\sim& \Big(1 + \star\Big)  \Big(1+ P \ln \Big(\frac{r}{r_0}\Big)\bigg)  \dot{\psi} \ du \wedge dr_1 \wedge r_1 d\phi_1 \wedge dr_2 \wedge r_2 d\phi_2 \ . 
\end{eqnarray}
It has been noted that \cite{PandoZayas:2002dso} this background leads to pp-wave upon 
setting $P=0$. As we have noted earlier, the constant $P$ is proportional to $M$.
Thus, setting $P$ to zero amounts to removing the fractional $D3$ branes,
there by restoring to the undeformed Klebanov-Witten background. This indicates
that the deformed background does not support pp-wave upon taking Penrose limits.
In appendix A, we consider an extensive study of Penrose limits along the 
remaining null geodesics. Some of these geometries become singular where as 
some other are smooth. Nevertheless none of these limits give rise to pp-wave solution.
However as we will see in the following, this is not the case upon considering
the non-Abelian T-duality. Thus, non-Abelian T-duality gives rise to new exactly
solvable backgrounds that are absent in the original type $IIB$ configuration.

In order to carry out the Penrose limit along appropriate null geodesics of the 
non-Abelian T-dual background, we 
will first rescale various quantities appropriately. Let us first consider
the wrap factor $H(r)$. Introducing the parameter $\tilde r$ via
$$\ln\tilde{r} = \ln r_0 - \frac{1}{4} - \frac{R^4}{2L^4} \ ,  $$
we can rewrite it as 
\begin{eqnarray} \label{re H}
H(r) = \frac{2L^4}{r^4} \ln\bigg(\frac{r}{\tilde{r}}\bigg) \ .  
\end{eqnarray}
We will now rescale the Minkowski coordinates $x^\mu \rightarrow L^2 x^\mu$ and 
the T-dual coordinates $v_{2,3} \rightarrow L^2 v_{2,3}$. In terms of these 
rescaled coordinates,  the T-dual metric \eqref{NATD-dual metric} becomes 
\begin{eqnarray} \label{rescaled NATD-dual metric}
&d\hat{s}^2_{\text{NATD}}=&   L^2 \Bigg[\frac{1}{\sqrt{2}} \frac{r^2}{{\sqrt{\ln\Big({\frac{r}{\tilde{r}}\Big)}}}}  \ \eta_{\mu\nu} dx^{\mu} dx^{\nu}+  \sqrt{2} \ \frac{\sqrt{\ln\Big({\frac{r}{\tilde{r}}\Big)}}}{r^2} \Big(dr^2 + \frac{1}{6} r^2  d\Omega_2^2  (\theta_1, \phi_1) \Big)\Bigg] \ \nonumber\\
&& + \ \frac{L^2}{A} \Bigg[6 \sqrt{2} \ \sqrt{\ln\Big({\frac{r}{\tilde{r}}\Big)}} \ v_2^2 \sigma_{\hat{3}}^2 +  \Big(6 \sqrt{2} \ \sqrt{\ln\Big({\frac{r}{\tilde{r}}\Big)}} +  \frac{81 \sqrt{2}}{\sqrt{\ln\Big({\frac{r}{\tilde{r}}\Big)}}} \ v_2^2 \Big) dv_2^2 \nonumber\\
&&+ \bigg(9 \sqrt{2} \ \sqrt{\ln\Big({\frac{r}{\tilde{r}}\Big)}} + \frac{9}{2\sqrt{2}} \  \frac{1}{\sqrt{\ln\Big({\frac{r}{\tilde{r}}\Big)}}} \ \Big(6v_3 + 2\sqrt{2}  \ln\Big(\frac{r}{r_0}\Big) \Big)^2 \bigg) dv_3^2 \nonumber\\
&& + \ \frac{27 \sqrt{2}}{ \sqrt{\ln\Big({\frac{r}{\tilde{r}}\Big)}}} \ \bigg(6v_3 + 2\sqrt{2}  \ln\Big(\frac{r}{r_0}\Big) \bigg)\ v_2 dv_2 dv_3 \Bigg] \ .
\end{eqnarray}
In the above, for easy reading we have introduced the notation
$$A = 4 \ln\Big({\frac{r}{\tilde{r}}\Big)} + \bigg(6v_3 + 2\sqrt{2} \ln\Big(\frac{r}{r_0}\Big) \bigg)^2 + 54 v_2^2 \ . $$
After the rescaling, the NS-NS two form $\hat{B}_2$ and the dilaton $\hat\Phi$ 
becomes
\begin{eqnarray} \label{rescaled NATD NS-NS sector}
&\hat{B}_2=&    \frac{L^2}{3} \ln\Big(\frac{r}{r_0}\Big) \sin\theta_1 d\theta_1 \wedge d\phi_1 + \frac{L^2}{A} \Bigg[3\sqrt{2} \bigg(6v_3 + 2\sqrt{2} \ln\Big(\frac{r}{r_0}\Big) \bigg) \ v_2 \ \sigma_{\hat{3}} \wedge dv_2 \nonumber\\
&& + \bigg(2\sqrt{2} \ \ln\Big({\frac{r}{\tilde{r}}\Big)} + \frac{1}{\sqrt{2}} \bigg(6v_3 + 2\sqrt{2} \ln\Big(\frac{r}{r_0}\Big) \bigg)^2\bigg)\ \sigma_{\hat{3}} \wedge dv_3\Bigg] \ , \nonumber\\
&e^{-2\hat{\Phi}}=&  \frac{L^6}{81 g_s^2} \ \sqrt{2} \ \sqrt{\ln\Big({\frac{r}{\tilde{r}}\Big)}} \ A \ .
\end{eqnarray}
Similarly, the field strengths in the RR sectors becomes
\begin{eqnarray} \label{rescaled NATD RR sector}
&\hat{F}_ 0=& - L^2 \ \frac{2\sqrt{2}}{9 g_s} \ ,   \nonumber\\
&\hat{F}_ 2=&   \frac{L^4}{162 \sqrt{2} \ g_s} \Bigg[\frac{2}{5} \ln\Big(\frac{r}{\tilde{r}}\Big) - \frac{1}{10} + 6\sqrt{2}  \Bigg(6v_3 + 2\sqrt{2} \ln\Big(\frac{r}{r_0}\Big) \Bigg)\Bigg] \sin\theta_1 d\theta_1 \wedge d\phi_1 \nonumber\\
&& + \ \frac{L^4}{g_s A} \frac{4}{3} \Bigg[-\Bigg(6v_3 + 2\sqrt{2} \ln\Big(\frac{r}{r_0}\Big)\Bigg) v_2 \ \sigma_{\hat{3}} \wedge dv_2 + 9 v_2^2 \ \sigma_{\hat{3}} \wedge dv_3\Bigg] \ , \nonumber\\
&\hat{F}_ 4=&  \frac{L^6}{g_s A}  \ v_2 \ \sin\theta_1 d\theta_1 \wedge d\phi_1 \wedge d\psi \wedge \Bigg[\Bigg(-\sqrt{2}  \Bigg(6v_3 + 2\sqrt{2} \ln\Big(\frac{r}{r_0}\Big) \Bigg) - \frac{1}{15} \ln\Big(\frac{r}{\tilde{r}}\Big)  \nonumber\\
&& \ + \frac{1}{60} \Bigg) v_2 dv_3 +  \Bigg(- \frac{4\sqrt{2}}{9} \ln\Big({\frac{r}{\tilde{r}}\Big)} +  \frac{1}{9} \bigg(\frac{1}{15}  \ln\Big(\frac{r}{\tilde{r}}\Big) - \frac{1}{60}\bigg) \bigg(6v_3 + 2\sqrt{2} \ln\Big(\frac{r}{r_0}\Big) \bigg)                 \nonumber\\
&&- 6\sqrt{2} v_2^2 \Bigg) dv_2 \Bigg] \ .
\end{eqnarray}

We will now consider the Penrose limits of the above T-dual background 
along appropriate null geodesics. Denoting the spacetime coordinates as
$\{x^\mu\}$, the geodesic equation is expressed as 
\begin{eqnarray} \label{geo}
 \frac{d^2 x^\mu}{du^2} + \Gamma_{\nu \rho}^\mu  \frac{dx^\nu}{du} \frac{dx^\rho}{du} = 0 \ .
\end{eqnarray}
Here we use $u$ to denote the affine parameter along the geodesic. We are 
interested to examine the motion along various isometry directions. Denote
$x^\lambda$ one such isometry direction. Thus, we need to set the velocity
and acceleration along any direction $x^\mu, \mu\neq\lambda$ zero:
\begin{eqnarray}
 \frac{dx^\mu}{du} = 0 = \frac{d^2 x^\mu}{du^2} \ , \ {\rm for} \  \mu \neq \lambda  \ .
\end{eqnarray}
Substituting the above in \eqref{geo}, we find that the geodesic equation
for motion along an isometry direction takes the simple form
\begin{eqnarray}
\partial^\mu g_{\lambda \lambda} = 0 \ .
\end{eqnarray}
In additon to the above condition, we need to impose $ds^2=0$ in order to 
obtain null geodesics for our purpose.

Let us now focus on various isometry directions of the T-dual geometry. A quick
inspection of the rescaled geometry \eqref{rescaled NATD-dual metric} indicates
that both $\psi$ and $\phi_1$ are isometry directions. Let us first consider the
motion along $\psi$ direction. The geodesics equation for this case is 
\begin{equation}\label{mupsi}
\partial_{\mu} g_{\psi \psi} = 0 \ . 
\end{equation}
From \eqref{rescaled NATD-dual metric} we note the relevant component of the metric:
\begin{eqnarray} \label{psi com}
g_{\psi \psi} = \frac{L^2}{A} \ 6 \sqrt{2} \ \sqrt{\ln\Big({\frac{r}{\tilde{r}}\Big)}} \ v_2^2  \ .
\end{eqnarray}
The metric component $g_{\psi\psi}$ depends upon $r,v_2$ and $v_3$.
For $\mu=r$, the geodesic condition \eqref{mupsi} leads to $v_2=0$. Similarly,
for $\mu = v_2, v_3$ we obtain $\{r = \tilde{r} , v_2 = 0\}$. However, for all
the above values, the metric component $g_{\psi\psi}$ in \eqref{psi com} vanishes
leading to singular geometries. In the following, we will no longer consider 
Penrose limits for such singular geometries.

We will now consider motion along the $\phi_1$-direction. Consider the metric 
component $g_{\phi_1\phi_1}$ along the $\phi_1$-direction:
\begin{eqnarray} \label{phi1 com}
g_{\phi_1 \phi_1} = L^2  \ \sqrt{\ln\Big({\frac{r}{\tilde{r}}\Big)}} \ \Bigg[\frac{1}{3\sqrt{2}} \sin^2\theta_1 + \frac{6 \sqrt{2}}{A}  \ v_2^2 \cos^2\theta_1\Bigg]   \ .
\end{eqnarray}
Let us analyse in detail the geodesic condition: 
$$ \partial_\mu g_{\phi_1\phi_1} = 0  \ . $$
For $\mu=r$, the above equation leads to $\{\theta_1 = (0, \pi), v_2 = 0\}$. For
$\mu=\theta_1$, we find $\{ r = \tilde{r}, \theta_1 = (0, \frac{\pi}{2}, \pi)\}$.
On the other hand, for $\mu = v_2, v_3$ we obtain 
$\{ r = \tilde{r}, \theta_1 = \frac{\pi}{2}, v_2 = 0\}$. The geodesic condition
is trivially satisfied for all other values of $\mu$. Consider the values  
$r= \tilde{r}$,  $\{\theta_1 = (0, \pi), v_2  = 0\}$. The metric component 
$g_{\phi_1\phi_1}$ in \eqref{phi1 com} vanishes for all these cases. Thus, they
lead to singular geometries. We will not consider these geodesics any more.

Finally, we will consider Penrose limit around $\theta_1 = \frac{\pi}{2}, v_2 = 0 = v_3$, keeping the $r$-coordinate constant, {\it i.e.}, $r= c$ for some constant 
$c \neq \tilde{r} \neq 0$. Consider the following expansion around this geodesic: 
\begin{eqnarray} 
&&x_i  = \frac{y_i}{L} \ ; \ i =1, 2, 3, \  r = c + \frac{x}{L} \ , \ \theta_1 = \frac{\pi}{2} + \frac{z}{L} \ ,  \ t = ax^+ \ , \nonumber\\
&&\phi_1 = bx^+ + \frac{x^-}{L^2} \ .
\end{eqnarray}
In addition, we do the rescaling of the coordinates $v_2$ and $v_3$ as $ \ v_2 \rightarrow \frac{v_2}{L} \ , \ v_3 \rightarrow \frac{v_3}{L} \ \   $
while keeping the $\psi$-coordinate unchanged. In the above, $a$ and $b$ are 
some constant parameters. The null geodesic condition relates the parameters
$a,b$ and $c$ as:
\begin{equation}\label{abc}
a^2 = \frac{b^2}{3c^2} \ln\Big(\frac{c}{\tilde{r}}\Big) \ . 
\end{equation}
Using the above expansion we consider the leading terms of the T-dual metric in 
the limit $L\rightarrow \infty$. We find 
\begin{eqnarray} 
&ds^2=&  \frac{1}{3\sqrt{2}}  \sqrt{\ln\Big({\frac{c}{\tilde{r}}\Big)}} \ 2b dx^+ dx^- +  \frac{1}{\sqrt{2}} \ \frac{c^2}{\sqrt{\ln\Big({\frac{c}{\tilde{r}}\Big)}}} \Big(dy_1^2 + dy_2^2 + dy_3^2\Big) +  \frac{\sqrt{2}}{c^2}  \sqrt{\ln\Big({\frac{c}{\tilde{r}}\Big)}} \ dx^2  \nonumber\\
 && + \frac{1}{3\sqrt{2}}  \sqrt{\ln\Big({\frac{c}{\tilde{r}}\Big)}} \ dz^2+ \frac{3}{\sqrt{2}} \ \frac{\sqrt{\ln\Big({\frac{c}{\tilde{r}}\Big)}}}{\ln\Big({\frac{c}{\tilde{r}}\Big)} + 2 \Big(\ln\Big(\frac{c}{r_0}\Big)\Big)^2 } \Big(dv_2^2 + v_2^2 d\psi^2\Big) +  \frac{9}{2 \sqrt{2}} \frac{1}{\sqrt{\ln\Big({\frac{c}{\tilde{r}}\Big)}}}\ dv_3^2 \nonumber\\
 && - \frac{b^2}{3\sqrt{2} c^2} \sqrt{\ln\Big({\frac{c}{\tilde{r}}\Big)}} \Bigg[\frac{x^2}{\ln\Big(\frac{c}{\tilde{r}}\Big)}  + x^2 + c^2 z^2\Bigg]   
(dx^+)^2 - L \ \frac{2b^2x}{3c} \ln\Big(\frac{c}{\tilde{r}}\Big) \ (dx^+)^2 \ .   
\end{eqnarray}
This contains a divergent term which can't be remove for any choice of the parameter $b$. Note that, from the null geodesic condition \eqref{abc}, 
setting $b=0$ is not allowed. Hence, motion along the isometric direction 
$\phi_1$ by keeping $r=$ constant does not lead to any smooth geometry.

We can repeat similar analysis for motion along the isometry direction $\psi$.
Recall that the ${\psi\psi}$ component of the T-dual metric is given as 
\begin{eqnarray} 
g_{\psi\psi} = \frac{L^2}{A} \ 6\sqrt{2} \ \sqrt{\ln\Big(\frac{r}{\tilde{r}}\Big)} \ v_2^2  \ . 
\end{eqnarray}
The null geodesic condition 
\begin{equation}
\partial_\mu g_{\psi\psi} = 0 \ ,
\end{equation}
for $\mu = r$ leads to $v_2 = 0$, where as, for $\mu = v_2, v_3$ we obtain 
$\{r = \tilde{r}, v_2 = 0 \}$. However, the metric component $g_{\psi\psi}$ 
vanishes for all these values. Thus, we do not have a regular geometry for 
any of the above geodesics.

To the end we will consider null geodesics for the motion of a particle 
carrying nonzero angluar momentum in the $(r,\phi_1)$ plane. We will
subject our analysis to a small neighbourhood of  
$\theta_1 = \frac{\pi}{2}$ and $v_2 = v_3 = 0$. 
Consider the Lagrangian for a massless particle moving along this geodesic:
\begin{eqnarray}\label{NATD lagrangian}
\mathcal{L} = \frac{1}{2} g_{\mu\nu} \dot{X}^\mu \dot{X}^\nu  \ .
\end{eqnarray}
Let $u$ be the affine parameter along the geodesic. The dots in the above
equation correspond to derivative with respect to $u$. Substituting the 
explicit expression for the background metric \eqref{rescaled NATD-dual metric} in the above Lagrangian we find
\begin{eqnarray}\label{NATD explict lag}
\mathcal{L} = \frac{L^2}{2} \Bigg(-  \frac{1}{\sqrt{2}} \frac{r^2}{\sqrt{\ln\Big({\frac{r}{\tilde{r}}\Big)}}} \ \dot{t}^2 + \frac{\sqrt{2}}{r^2} \sqrt{\ln\Big({\frac{r}{\tilde{r}}\Big)}} \ \dot{r}^2 +  \frac{1}{3\sqrt{2}}  \sqrt{\ln\Big({\frac{r}{\tilde{r}}\Big)}} \ \dot{\phi}_1^2\Bigg) \ .
\end{eqnarray}

We will now obtain the conserved quantities corresponding to the above 
system. Note that, the Lagrangian \eqref{NATD explict lag} does not 
depend on the generalized coordinates $t$ and $\phi_1$ explicitly. 
Denoting $- E L^2$ to be the conserved momentum associated with $t$,
we find 
\begin{eqnarray} \label{NATD t}
E = - \frac{1}{L^2}\frac{\partial \mathcal{L}}{\partial\dot{t}} =   \frac{r^2}{\sqrt{2 \ln\Big({\frac{r}{\tilde{r}}\Big)}}}   \ \dot{t}   \ .
\end{eqnarray}
Similarly, let $- J L^2 $  be the conserved momentum associated with the
generalized coordinate $\phi_1$. We find
\begin{eqnarray} \label{NATD phi-1}
J = - \frac{1}{L^2 }\frac{\partial \mathcal{L}}{\partial\dot{\phi}_1} = - \frac{1}{3\sqrt{2}}  \sqrt{\ln\Big({\frac{r}{\tilde{r}}\Big)}} \ \dot{\phi}_1  \ .
\end{eqnarray}
In addition, we will require the geodesic to be null. This gives rise to the 
condition:
\begin{eqnarray}
\dot{r}^2 +  \frac{3r^2}{\ln\Big({\frac{r}{\tilde{r}}\Big)}} J^2 = E^2  \ .
\end{eqnarray}

We will now concentrate on obtaining the Penrose limit for a null geodesic 
carrying angular momentum $J$ around $x_i  = 0 , i =1, 2, 3$ , $\theta_1 = \frac{\pi}{2}$ and $v_2 = v_3 = 0$. We redefine the coordinates as 

\begin{eqnarray} \label{NATD redefine}
x_i  = \frac{y_i}{L} \ ; \ i =1, 2, 3, \ \theta_1 = \frac{\pi}{2} + \frac{z}{L} \ , \ v_2 \rightarrow \frac{v_2}{L} \ , \ v_3 \rightarrow \frac{v_3}{L} \ .
\end{eqnarray}
We keep the $\psi$-coordinate unchanged, and redefine the string coupling as 
$g_s = L^3 \ \tilde{g}_s$, in order to keep the dilaton finite at the Penrose 
limit. Finally, we will consider the following expansion in the limit $L \rightarrow \infty$:
\begin{eqnarray} \label{NATD expansion}
dt = c_1 du , \ dr = c_2 du + c_3 \frac{dw}{L}  \ , \ d\phi_1 = c_4 du + c_5 \frac{dw}{L} + c_6 \frac{dv}{L^2}  \  .
\end{eqnarray}
We need to determine the coefficients $c_i$. Requiring the geodesic to be null
determines the values of the coefficients $c_1, c_2$ and $c_4$ as follows:
\begin{eqnarray} \label{NATD c-values}
c_1 &=& \frac{E \sqrt{2}}{r^2} \ \sqrt{\ln\Big({\frac{r}{\tilde{r}}\Big)}}   \ , \cr
 \ c_2 &=& \Bigg[E^2 -  \frac{3r^2}{\ln\Big({\frac{r}{\tilde{r}}\Big)}} J^2 \Bigg]^{\frac{1}{2}}  , \cr 
 \ c_4 &=& -  \frac{3\sqrt{2}}{\sqrt{\ln\Big({\frac{r}{\tilde{r}}\Big)}}}  \ J    \  .
\end{eqnarray}

We now substitute the expansion \eqref{NATD expansion} in the T-dual metric
\eqref{rescaled NATD-dual metric} and retain the leading terms. Apriory, this 
metric will contain divergent terms of order $L$ and $L^2$. Imposing the null 
geodesic condition automatically cancels the $\mathcal{O}(L^2)$ terms. It can 
be easily verified that the $\mathcal{O}(L)$ term is removed upon setting 
\begin{equation}
c_2 c_3 + \frac{r^2}{6} c_4 c_5 = 0 \ .
\end{equation}
{Using the value of $c_2$ and $c_4$ from \eqref{NATD c-values} in the above 
equation we can express the coefficient $c_3$ in terms $c_5$ as }
\begin{eqnarray}
c_3 = \Bigg[E^2 -  \frac{3r^2}{\ln\Big({\frac{r}{\tilde{r}}\Big)}} J^2 \Bigg]^{- \frac{1}{2}}  \frac{r^2}{\sqrt{2 \ln\Big({\frac{r}{\tilde{r}}\Big)}}}  \ J  \  c_5  \ . 
\end{eqnarray}
We will see later that the coefficient $c_5$ can be determined by requiring
the background fields to satisfy the Einstein's equations. Finally, we need
to determine the coefficient $c_6$. This is easily obtained by upon setting 
appropriate normalization for the cross term $du dv$ in the metric. We find
\begin{equation} c_6 = - \frac{1}{J} \ . 
\end{equation}

Substituting the above results in \eqref{rescaled NATD-dual metric}, and taking 
the limit $L\rightarrow\infty$, we find the pp-wave metric of the form
\begin{eqnarray} \label{NATD penrose limit metric}
&ds^2=&  2 du dv + \frac{1}{\sqrt{2}} \ \frac{r^2}{\sqrt{\ln\Big({\frac{r}{\tilde{r}}\Big)}}} \Big(dy_1^2 + dy_2^2 + dy_3^2\Big) + \sqrt{\ln\Big({\frac{r}{\tilde{r}}\Big)}} \ \bigg(\frac{c_3^2 \sqrt{2}}{r^2} + \frac{c_5^2}{3\sqrt{2}}\bigg) dw^2 \nonumber\\
&& + \ \frac{1}{3\sqrt{2}}  \sqrt{\ln\Big({\frac{r}{\tilde{r}}\Big)}} \ dz^2 + \frac{3}{\sqrt{2}} \ \frac{\sqrt{\ln\Big({\frac{r}{\tilde{r}}\Big)}}}{\ln\Big({\frac{r}{\tilde{r}}\Big)} + 2 \Big(\ln\Big(\frac{r}{r_0}\Big)\Big)^2 } \Big(dv_2^2 + v_2^2 d\psi^2\Big) \nonumber\\
&&\ +  \frac{9}{2 \sqrt{2}} \frac{1}{\sqrt{\ln\Big({\frac{r}{\tilde{r}}\Big)}}}\ dv_3^2  -  \frac{3\sqrt{2}}{\sqrt{\ln\Big({\frac{r}{\tilde{r}}\Big)}}} J^2 z^2  \ du^2 \ .     
\end{eqnarray}
We will subsequently show that this is indeed a pp-wave solution by rewriting
it in the standard Brinkmann form. Let us now consider the Penrose limit for 
the remaining background fields. In this limit, the NS-NS two-form field and 
dilaton takes the form 
\begin{eqnarray} \label{NS-NS}
&\hat{B}_2 =&   \frac{c_5}{3} \ln\Big({\frac{r}{r_0}\Big)} \ dz \wedge dw - 3 \ \frac{\ln\Big(\frac{r}{r_0}\Big)}{\ln\Big({\frac{r}{\tilde{r}}\Big)} + 2 \Big(\ln\Big(\frac{r}{r_0}\Big)\Big)^2}  \ v_2 dv_2 \wedge d\psi + \frac{3J z}{\sqrt{\ln\Big(\frac{r}{\tilde{r}}\Big)}} \ du \wedge dv_3  \ ,                                                                      \nonumber\\
&e^{-2\hat{\Phi}}=&  \frac{4 \sqrt{2} }{81 \ \tilde{g}_s^2}  \ \sqrt{\ln\Big({\frac{r}{\tilde{r}}\Big)}}  \ \Bigg[\ln\Big({\frac{r}{\tilde{r}}\Big)} + 2 \Big(\ln\Big(\frac{r}{r_0}\Big)\Big)^2 \Bigg] \ .
\end{eqnarray}
The field strengths corresponding to the RR sector are given by
\begin{eqnarray} \label{penrose limit RR}
 &\hat{F}_0=&  0 \ ,  \nonumber\\
 &\hat{F}_2=&  \frac{J}{54 \tilde{g}_s}  \frac{1}{\sqrt{\ln\Big({\frac{r}{\tilde{r}}\Big)}}} \Bigg[\frac{2}{5} \ln\Big(\frac{r}{\tilde{r}}\Big) + 24 \ln\Big(\frac{r}{r_0}\Big) - \frac{1}{10}\Bigg] du \wedge dz  \ ,      \nonumber\\
 &\hat{F}_4=&   \frac{2J}{3\tilde{g} _s \Big(\ln\Big({\frac{r}{\tilde{r}}\Big)} + 2 \Big(\ln\Big(\frac{r}{r_0}\Big)\Big)^2\Big)}  \frac{1}{\sqrt{\ln\Big({\frac{r}{\tilde{r}}\Big)}}} \ v_2  \Bigg[- \ln\Big(\frac{r}{\tilde{r}}\Big) + \frac{1}{2} \ln\Big(\frac{r}{r_0}\Big) \Big(\frac{1}{15} \ln\Big(\frac{r}{\tilde{r}}\Big) \nonumber\\
 &&-\frac{1}{60}\Big)  \Bigg] du \wedge dz \wedge d\psi \wedge dv_2 \ . 
\end{eqnarray}

For later use, we will also compute the field strength $\hat H_3$ corresponding
to the NS-NS two-form $\hat B_2$:
\begin{eqnarray} \label{h3}
&\hat{H}_3=&   \frac{1}{3} \Bigg[E^2 -  \frac{3r^2}{\ln\Big({\frac{r}{\tilde{r}}\Big)}} J^2 \Bigg]^{\frac{1}{2}} \Bigg[c_5^{\prime} \ln\Big({\frac{r}{r_0}\Big)} + \frac{c_5}{r}\Bigg] du \wedge dz \wedge dw  - 3v_2  \Bigg[E^2 -  \frac{3r^2}{\ln\Big({\frac{r}{\tilde{r}}\Big)}} J^2 \Bigg]^{\frac{1}{2}} \nonumber\\
&&\frac{\ln\Big(\frac{r}{\tilde{r}}\Big) -\ln\Big(\frac{r}{r_0}\Big) - 2\Big(\ln\Big(\frac{r}{r_0}\Big)\Big)^2  }{r \Big(\ln\Big({\frac{r}{\tilde{r}}\Big)} + 2 \Big(\ln\Big(\frac{r}{r_0}\Big)\Big)^2\Big)^2} \ du \wedge dv_2 \wedge d\psi + \frac{3J}{\sqrt{\ln\Big(\frac{r}{\tilde{r}}\Big)}} \ du \wedge dv_3 \wedge dz \  . \nonumber\\
\end{eqnarray}
Note that, in obtaining the above, we have used $dr=c_2 du$ where the expression
for the coefficient $c_2$ is given by \eqref{NATD c-values}.

As pointed out earlier, the metric obtained in \eqref{NATD penrose limit metric} 
is not in the standard {{Brinkmann}} form \cite{blau.m}. A formalism has been
developed in \cite{Itsios:2017nou} in order to transform the line element to the Brinkmann form. Following \cite{Itsios:2017nou} consider a line element of the 
form
\begin{eqnarray} \label{NATD line element}
ds^2 = 2 dudv + \sum_i A_i (u) \ dx_i ^2 \ .
\end{eqnarray}
Now, replace the coordinates $x_i$ and $v$ as 
\begin{eqnarray} \label{transformation}
x_i \rightarrow \frac{x_i}{\sqrt{A_i}} \ ,  \ v \rightarrow v + \frac{1}{4} \sum_i \frac{\dot{A_i}}{A_i} x_i^2 \ .
\end{eqnarray}
The line element in  \eqref{NATD line element} now takes the Brinkmann form 
\begin{eqnarray} \label{Brinkmann line element}
ds^2 = 2 dudv + \sum_i dx_i ^2 + \Big(\sum_i F_i (u) x_i ^2\Big) \ du^2  \ ,
\end{eqnarray}
with the functions $F_i$ being
\begin{eqnarray} \label{Fi definitions }
F_i = \frac{1}{4} \frac{\dot{A_i^2}}{A_i^2} \ + \ \frac{1}{2} \frac{d}{du} \Big(\frac{\dot{A_i}}{A_i}\Big) \ .
\end{eqnarray}

For the case of our pp-wave metric \eqref{NATD penrose limit metric} we have
\begin{eqnarray} \label{Ai}
&&A_{y_{1}} = A_{y_{2}} = A_{y_{3}} = \frac{1}{\sqrt{2}} \ \frac{r^2}{\sqrt{\ln\Big({\frac{r}{\tilde{r}}\Big)}}}  \ ,  \ A_w =  \sqrt{\ln\Big({\frac{r}{\tilde{r}}\Big)}} \ \bigg(\frac{c_3^2 \sqrt{2}}{r^2} + \frac{c_5^2}{3\sqrt{2}}\bigg) \  , \nonumber\\
&&A_z = \frac{1}{3\sqrt{2}}  \sqrt{\ln\Big({\frac{r}{\tilde{r}}\Big)}} \ ,\  A_{v_2} = \frac{3}{\sqrt{2}} \ \frac{\sqrt{\ln\Big({\frac{r}{\tilde{r}}\Big)}}}{\ln\Big({\frac{r}{\tilde{r}}\Big)} + 2 \Big(\ln\Big(\frac{r}{r_0}\Big)\Big)^2 } \ , \ A_{v_3} = \frac{9}{2 \sqrt{2}} \frac{1}{\sqrt{\ln\Big({\frac{r}{\tilde{r}}\Big)}}}    \ .  \nonumber\\
\end{eqnarray}
Hence after making the following replacement
\begin{eqnarray} \label{coordinate transformation}
&&y_1  \rightarrow \frac{y_1}{\sqrt{A_{y_1}}} \ , \ y_2  \rightarrow \frac{y_2}{\sqrt{A_{y_2}}} \ ,  \ y_3  \rightarrow \frac{y_3}{\sqrt{A_{y_3}}} \ ,   \ w  \rightarrow \frac{w}{\sqrt{A_{w}}} \ ,   \ z  \rightarrow \frac{z}{\sqrt{A_{z}}} \ ,  \nonumber\\
&& v_2 \rightarrow \frac{v_2}{\sqrt{A_{v_2}}} \ , \ v_3 \rightarrow \frac{v_3}{\sqrt{A_{v_3}}} \  \ \text{and}  \nonumber\\
&& v \rightarrow v + \frac{1}{4} \Bigg[\frac{\dot{A}_{y_{1}}}{A_{y_{1}}} \ y_1 ^2  + \frac{\dot{A}_{y_{2}}}{A_{y_{2}}} \ y_2 ^2 + \frac{\dot{A}_{y_{3}}}{A_{y_{3}}} \ y_3 ^2 + \frac{\dot{A}_w}{A_w} \ w ^2 + \frac{\dot{A}_z}{A_z} \ z ^2 + \frac{\dot{A}_{v_{2}}}{A_{v_{2}}} \ v_2^2  + \frac{\dot{A}_{v_{3}}}{A_{v_{3}}} \ v_3^2\Bigg] \ , \nonumber\\
\end{eqnarray}
we find
\begin{eqnarray} \label{metric brinkmann}
&ds^2 =&  2 dudv + dy_1^2 + dy_2^2 + dy_3^2 + dw^2 + dz^2 + dv_2^2 + v_2^2 \ d\psi^2 + dv_3^2  \nonumber\\
&&+ \Bigg[F_{y_{1}}  y_1^2 + F_{y_{2}}  y_2^2 + F_{y_{3}}  y_3^2 + F_w w^2 +  F_z z^2 + F_{v_{2}} v_2^2 + F_{v_{3}}  v_3^2  - \frac{3\sqrt{2}}{\sqrt{\ln\Big({\frac{r}{\tilde{r}}\Big)}}} \ J^2 z^2  \Bigg]du^2  \ , \nonumber\\
\end{eqnarray}
where the functions $F_i$ can be read from the expression \eqref{Fi definitions }. 

We will now express the background fields in the Brinkmann form. The dilaton
$\hat\Phi$ and the NS-NS three form flux $\hat H_3$ are given as 
\begin{eqnarray} \label{NS-NS Brinkmann}
&e^{-2\hat{\Phi}}=&  \frac{4 \sqrt{2} }{81 \ \tilde{g}_s^2}  \ \sqrt{\ln\Big({\frac{r}{\tilde{r}}\Big)}}  \ \Bigg[\ln\Big({\frac{r}{\tilde{r}}\Big)} + 2 \Big(\ln\Big(\frac{r}{r_0}\Big)\Big)^2 \Bigg] \ , \nonumber\\
 &\hat{H}_3=&   \frac{2^{\frac{1}{4}}}{\sqrt{3}}  \Bigg[E^2 -  \frac{3r^2}{\ln\Big({\frac{r}{\tilde{r}}\Big)}} J^2 \Bigg]^{\frac{1}{2}} \Bigg[c_5^{\prime} \ln\Big({\frac{r}{r_0}\Big)} + \frac{c_5}{r}\Bigg] \Bigg[\frac{c_3^2 \sqrt{2}}{r^2} + \frac{c_5^2}{3\sqrt{2}} \Bigg]^ {- \frac{1}{2}}   \frac{1}{\sqrt{\ln\Big(\frac{r}{\tilde{r}}\Big)}}  \nonumber\\
 &&\ du \wedge dz \wedge dw - \ \sqrt{2} v_2  \Bigg[E^2 -  \frac{3r^2}{\ln\Big({\frac{r}{\tilde{r}}\Big)}} J^2 \Bigg]^{\frac{1}{2}} \ \frac{\ln\Big(\frac{r}{\tilde{r}}\Big) -\ln\Big(\frac{r}{r_0}\Big) - 2\Big(\ln\Big(\frac{r}{r_0}\Big)\Big)^2  }{r \Big(\ln\Big({\frac{r}{\tilde{r}}\Big)} + 2 \Big(\ln\Big(\frac{r}{r_0}\Big)\Big)^2\Big)}   \nonumber\\
 &&\ \frac{1}{\sqrt{\ln\Big(\frac{r}{\tilde{r}}\Big)}} \ du \wedge dv_2 \wedge d\psi  + \frac{2 \sqrt{3} \ J}{\sqrt{\ln\Big(\frac{r}{\tilde{r}}\Big)}} \ du \wedge dv_3 \wedge dz \  . 
\end{eqnarray}

Similarly, the expressions for the RR field strengths are found to be of the form
\begin{eqnarray} \label{Brinkmann RR}
 &\hat{F}_0=&  0 \ ,  \nonumber\\
 &\hat{F}_2=&   \frac{2^{(- \frac{3}{4})}}{9\sqrt{3} \ \tilde{g}_s} \frac{J}{{\Big(\ln\Big({\frac{r}{\tilde{r}}\Big)\Big)^{\frac{3}{4}}}}} \Bigg[\frac{2}{5} \ln\Big(\frac{r}{\tilde{r}}\Big) + 24 \ln\Big(\frac{r}{r_0}\Big) - \frac{1}{10}\Bigg] du \wedge dz  \ ,       \nonumber\\
 &\hat{F}_4=&   \frac{2^{ \frac{7}{4}}}{3\sqrt{3} \ \tilde{g}_s} \frac{J}{{\Big(\ln\Big({\frac{r}{\tilde{r}}\Big)\Big)^{\frac{5}{4}}}}} \ v_2 \Bigg[-\ln\Big(\frac{r}{\tilde{r}}\Big) + \frac{1}{2} \ln\Big(\frac{r}{r_0}\Big) \bigg(\frac{1}{15} \ln\Big(\frac{r}{\tilde{r}}\Big) \nonumber\\
 &&-\frac{1}{60}\bigg)  \Bigg] du \wedge dz \wedge d\psi \wedge dv_2 \ . 
\end{eqnarray}

We will now verify that these fields indeed satisfy the Bianchi identities 
and the gauge field equation of motion. A quick inspection of the background 
fields in \eqref{NS-NS Brinkmann} -\eqref{Brinkmann RR} shows that the 
Bianchi identities 
\begin{eqnarray} \label{Bianchi 2a natd}
&&d\hat{H}_3 = 0 \ , \  d\hat{F}_2 = \hat{F}_0 \hat{H}_3 \ , \ d\hat{F}_4 = \hat{H}_3 \wedge \hat{F}_2 \ 
\end{eqnarray}
hold. The field strengths $\hat H_3, \hat F_2$ and $\hat F_4$ are all closed
and both $\hat F_0$ as well as $\hat{H}_3 \wedge \hat{F}_2$ are indeed zero.

Let us now inspect the type-$IIA$ supergravity equations for the gauge fields
\begin{eqnarray} \label{Bianchi 2a natd1}
%&&d\hat{H}_3 = 0 \ , \  d\hat{F}_2 = \hat{F}_0 \hat{H}_3 \ , \ d\hat{F}_4 = \hat{H}_3 \wedge \hat{F}_2 \ , \nonumber\\
&&d\Big(e^{-2\hat{\Phi}} \star \hat{H}_3\Big) - \hat{F}_2 \wedge \star \hat{F}_4 - \frac{1}{2}  \hat{F}_4 \wedge \hat{F}_4 = \hat{F}_0 \star \hat{F}_2 \ , \nonumber\\
&&d\star \hat{F}_2 + \hat{H}_3 \wedge \star \hat{F}_4 = 0 \ , \nonumber\\
&&d\star \hat{F}_4 + \hat{H}_3 \wedge \hat{F}_4 = 0 \ .
\end{eqnarray}
The Hodge duals for the above background fields are 
\begin{eqnarray} \label{Hodge dual}
&\star \ \hat{H}_3=&   \frac{1}{7 !} \ du \wedge dy_1 \wedge dy_2 \wedge dy_3 \Bigg[ \frac{2^{\frac{1}{4}}}{\sqrt{3}} \ v_2  \bigg(E^2 -  \frac{3r^2}{\ln\Big({\frac{r}{\tilde{r}}\Big)}} J^2 \bigg)^{\frac{1}{2}} \bigg(c_5^{\prime} \ln\Big({\frac{r}{r_0}\Big)} + \frac{c_5}{r}\bigg)  \nonumber\\
&&\bigg(\frac{c_3^2 \sqrt{2}}{r^2} + \frac{c_5^2}{3\sqrt{2}} \bigg)^ {- \frac{1}{2}}  \frac{1}{\sqrt{\ln\Big(\frac{r}{\tilde{r}}\Big)}}  \ dv_2 \wedge d\psi \wedge dv_3   \nonumber\\ 
&&-  \sqrt{2}   \bigg(E^2 -  \frac{3r^2}{\ln\Big({\frac{r}{\tilde{r}}\Big)}} J^2 \bigg)^{\frac{1}{2}} \ \frac{\ln\Big(\frac{r}{\tilde{r}}\Big) -\ln\Big(\frac{r}{r_0}\Big) - 2\Big(\ln\Big(\frac{r}{r_0}\Big)\Big)^2  }{r \Big(\ln\Big({\frac{r}{\tilde{r}}\Big)} + 2 \Big(\ln\Big(\frac{r}{r_0}\Big)\Big)^2\Big)^2}                \nonumber\\
&&\ \frac{1}{\sqrt{\ln\Big(\frac{r}{\tilde{r}}\Big)}}  \ dz \wedge dw \wedge dv_3  +  \ v_2   \ \frac{2 \sqrt{3} \ J}{\sqrt{\ln\Big(\frac{r}{\tilde{r}}\Big)}} \ dw \wedge dv_2 \wedge d\psi \Bigg] \ , \nonumber\\ 
&& \nonumber\\
&\star \ \hat{F}_2 =& \frac{1}{8!} \ \frac{2^{(- \frac{3}{4})}}{9\sqrt{3} \ \tilde{g}_s} \frac{J}{{\Big(\ln\Big({\frac{r}{\tilde{r}}\Big)\Big)^{\frac{3}{4}}}}} \ v_2 \Bigg[\frac{2}{5} \ln\Big(\frac{r}{\tilde{r}}\Big) + 24 \ln\Big(\frac{r}{r_0}\Big) 
- \frac{1}{10}\Bigg]    du \wedge dy_1 \wedge dy_2 \wedge dy_3 \nonumber\\
&&\wedge \ dw \wedge d\psi \wedge dv_2 \wedge dv_3    \ ,                \nonumber\\
&& \nonumber\\ 
&\star \ \hat{F}_4=&  \frac{1}{6!} \  \frac{2^{ \frac{7}{4}}}{3\sqrt{3} \ \tilde{g}_s} \frac{J}{{\Big(\ln\Big({\frac{r}{\tilde{r}}\Big)\Big)^{\frac{5}{4}}}}}  \  \Bigg[- \ln\Big(\frac{r}{\tilde{r}}\Big) + \frac{1}{2} \ln\Big(\frac{r}{r_0}\Big) \bigg(\frac{1}{15} \ln\Big(\frac{r}{\tilde{r}}\Big) -\frac{1}{60}\bigg)  \Bigg] du \wedge dy_1  \nonumber\\
&&\wedge dy_2 \wedge dy_3 \wedge \ dw \wedge dv_3  \ .  
\end{eqnarray}

In deriving the above, we have used 
$$
{\rm det}(g_{pp})= -v_2^2 \ , \ g^{vv} = - g_{uu} \ ,    g^{uv} = g^{vu} = 1 \ , \ g^{ij} = \delta^{ij} \Big(\text{only} \ g^{\psi \psi} = \frac{1}{v_2^2}\Big) \ . 
$$

It is straightforward to see that, the background fields \eqref{NS-NS Brinkmann}-\eqref{Brinkmann RR} together with \eqref{Hodge dual} indeed satisfy the gauge field equations for type-$IIA$ supergravity. Both $\star\hat H_3$ as well as 
$e^{-2\hat\Phi}\star\hat H_3$ are closed. Also, $\hat{F}_2 \wedge \star \hat{F}_4$
and $\hat{F}_4 \wedge \hat{F}_4$ vanish identically. Thus, the first of the 
equations in \eqref{Bianchi 2a natd1} is satisfied. Similarly, both $\star\hat F_2$
and $\star\hat F_4$ are exact forms. In addition, $\hat{H}_3 \wedge \hat{F}_4$
and $\hat{H}_3 \wedge \star \hat{F}_4$ vanish as well. Thus, the last two 
equations in \eqref{Bianchi 2a natd1} are also satisfied. 

It is interesting to note that the gauge field equations as well as the Bianchi
identities hold irrespective of the value of the coefficient $c_5$. However, as 
we will see in the following, this is not the case with the Einstein's equations. 
For type-$IIA$ supergravity, the Einstein's equations are given as
\begin{eqnarray} \label{Einstein eq natd}
\hat{R}_{\mu\nu} + 2D_{\mu}D_{\nu}\hat{\Phi} = \frac{1}{4} \hat{H}_{\mu\nu}^2 + e^{2\hat{\Phi}} \Bigg[\frac{1}{2} (\hat{F}_2^2)_{\mu\nu} + \frac{1}{12} (\hat{F}_4^2)_{\mu\nu}- \frac{1}{4} g_{\mu\nu} \Big(\hat{F}_{0}^2 + \frac{1}{2} \hat{F}_{2}^2 + \frac{1}{4!}\hat{F}_{4}^2 \Big)\Bigg]  \ . 
\end{eqnarray}
Similarly, the dilation equations are 
\begin{eqnarray} \label{R eq natd}
\hat{R} + 4D^2\hat{\Phi} - 4(\partial\hat{\Phi})^2 - \frac{1}{12}\hat{H}^2=0 \ .
\end{eqnarray}
In the appendix we have analysed these equations in detail. We find that the 
equation of motion for dilation \eqref{R eq natd} holds automatically. In 
addition, we observe that
the Einstein's equations \eqref{Einstein eq natd} are trivially satisfied for 
all values of $\mu,\nu$ except for $\mu=\nu=u$. In this case we have the 
nontrivial condition
\begin{eqnarray} \label{Einstein uu}
\hat{R}_{uu} + 2D_{u}D_{u}\hat{\Phi} = \frac{1}{4} \hat{H}_{uu}^2 + e^{2\hat{\Phi}} \Bigg[\frac{1}{2} (\hat{F}_2^2)_{uu} + \frac{1}{12} (\hat{F}_4^2)_{uu} \Bigg]  \ .
\end{eqnarray}
This equation involves the undetermined coefficient $c_5$. We can solve this
equation to determine the expression for the coefficient $c_5$. The analysis
in appendix B outlines the steps to determine it.

\section{Supersymmetry of pp-wave}

The supersymmatry analysis of non-Abelian T-dual backgrounds have been 
studied extensively \cite{Lozano:2011kb,Itsios:2012dc,Itsios:2012zv,Hassan:1999bv,Jeong:2013jfc,Kelekci:2014ima}. Unlike the $AdS_5 \times S^5$ case, the non-Abelian 
T-dual of Klebanov-Witten as well as the Klebanov-Tseytlin background preserves 
all the supersymmetries of the original background. This is because in the later
two cases,  the Killing spinor of the original background does not carry any
$SU(2)$ charge of the isometry group used for non-Abelian T-dualization
\cite{Lozano:2011kb,Kelekci:2014ima}. In
this context, it is worth investigating whether the pp-wave we obtained in 
the above preserves any supersymmetry.

In order to analyse this, we will first introduce the Brinkmann coordinates 
$X_i$ such that
\begin{eqnarray} \label{x}
dy_i^2 = \big(dX^i\big)^2 \ ; \ i =1,2,3 \ ,    \ w = X^4 \ , \ z = X^5\ , \cr
 \ dv_2^2 + v_2^2\ d\psi^2 = \big(dX^6\big)^2 + \big(dX^7\big)^2 \ ,  \ v_3 = X^8 \ .  \nonumber\\
\end{eqnarray}
In these coordinates the pp-wave background \eqref{metric brinkmann}-\eqref{Brinkmann RR} reads as
\begin{eqnarray} \label{in X coordinates} 
&ds^2 =& 2 dudv + \sum_{i=1}^{8} dX_i^2 + \mathcal{H} \ du^2 \ ,  \nonumber\\
&\hat{\Phi}=& \Phi(u) \ , \nonumber\\
&\hat{H}_3 =& f_1(u) \ du \wedge dX^5 \wedge dX^4 -  f_2(u) \ du \wedge dX^6 \wedge dX^7 + f_3(u) \ du \wedge dX^8 \wedge dX^5 \ ,                       \nonumber\\
&\hat{F}_2 =& f_4(u) \ du \wedge dX^5 \ ,                      \nonumber\\
&\hat{F}_4 =&  f_5(u) \ du \wedge dX^5 \wedge dX^7 \wedge dX^6 \ ,    
\end{eqnarray}
where we have introduced the notation
\begin{eqnarray} \label{Hf} 
&\mathcal{H} =& F_{ij} X^i X^j = \Bigg[F_{X^{1}}  (X^1)^2 + F_{X^{2}}  (X^2)^2+ F_{X^{3}}  (X^3)^2 + F_{X^{4}}  (X^4)^2 \cr
&+&  \Big(F_{X^{5}}  - \frac{3\sqrt{2}}{\sqrt{\ln\Big({\frac{r}{\tilde{r}}\Big)}}} \ J^2\Big)   (X^5)^2 +F_{X^{6}}  (X^6)^2 + F_{X^{7}}  (X^7)^2 + F_{X^{8}}  (X^8)^2 \Bigg] \ ,  \nonumber\\
&f_1(u) =&  \frac{2^{\frac{1}{4}}}{\sqrt{3}}  \Bigg[E^2 -  \frac{3r^2}{\ln\Big({\frac{r}{\tilde{r}}\Big)}} J^2 \Bigg]^{\frac{1}{2}} \Bigg[c_5^{\prime} \ln\Big({\frac{r}{r_0}\Big)} + \frac{c_5}{r}\Bigg] \Bigg[\frac{c_3^2 \sqrt{2}}{r^2} + \frac{c_5^2}{3\sqrt{2}} \Bigg]^ {- \frac{1}{2}}   \frac{1}{\sqrt{\ln\Big(\frac{r}{\tilde{r}}\Big)}}   \ ,                       \nonumber\\
&f_2(u) =& \sqrt{2}  \Bigg[E^2 -  \frac{3r^2}{\ln\Big({\frac{r}{\tilde{r}}\Big)}} J^2 \Bigg]^{\frac{1}{2}} \ \frac{\ln\Big(\frac{r}{\tilde{r}}\Big) -\ln\Big(\frac{r}{r_0}\Big) - 2\Big(\ln\Big(\frac{r}{r_0}\Big)\Big)^2  }{r \Big(\ln\Big({\frac{r}{\tilde{r}}\Big)} + 2 \Big(\ln\Big(\frac{r}{r_0}\Big)\Big)^2\Big)}   \frac{1}{\sqrt{\ln\Big(\frac{r}{\tilde{r}}\Big)}}  \ , \nonumber\\
&f_3(u) =& \frac{2 \sqrt{3} \ J}{\sqrt{\ln\Big(\frac{r}{\tilde{r}}\Big)}} \ ,   \nonumber\\
&f_4(u) =& \frac{2^{(- \frac{3}{4})}}{9\sqrt{3} \ \tilde{g}_s} \frac{J}{{\Big(\ln\Big({\frac{r}{\tilde{r}}\Big)\Big)^{\frac{3}{4}}}}} \Bigg[\frac{2}{5} \ln\Big(\frac{r}{\tilde{r}}\Big) + 24 \ln\Big(\frac{r}{r_0}\Big) - \frac{1}{10}\Bigg] \ ,     \nonumber\\
&f_5(u) =& \frac{2^{ \frac{7}{4}}}{3\sqrt{3} \ \tilde{g}_s} \frac{J}{{\Big(\ln\Big({\frac{r}{\tilde{r}}\Big)\Big)^{\frac{5}{4}}}}}  \Bigg[-\ln\Big(\frac{r}{\tilde{r}}\Big) + \frac{1}{2} \ln\Big(\frac{r}{r_0}\Big) \bigg(\frac{1}{15} \ln\Big(\frac{r}{\tilde{r}}\Big) -\frac{1}{60}\bigg)  \Bigg] \ . 
\end{eqnarray}
The functions $F_{ij}$ are defined by
 \begin{eqnarray} \label{Fij}
 &&F_{11} = F_{22} = F_{33} = F_ {X^i} \ ; \ i = 1,2,3 \ ,   \nonumber\\
 &&F_{44} = F_ {X^4} \ , \ F_{55} = F_ {X^5} - \frac{3\sqrt{2}}{\sqrt{\ln\Big({\frac{r}{\tilde{r}}\Big)}}} \ J^2 \ , \ F_{66} = F_ {X^6} \ , \ F_{77} = F_ {X^7} \ , \  F_{88} = F_ {X^8} \ . 
\end{eqnarray}

Now we introduce the frame $\{e^a\}$ as % for the pp-wave \eqref{in X coordinates} as follows 
\begin{eqnarray} \label{frame}
e^- = du \ , \ e^+ =  dv + \frac{1}{2} \mathcal{H} du \ ,  \  e^i = dX^i \ , 
\end{eqnarray}
such that the pp-wave metric \eqref{in X coordinates} can be written as 
\begin{eqnarray} 
ds^2 = 2e^+e^-  + \sum_{i=1}^8 \big(e^i\big)^2 = \eta_{ab} e^a e^b \ , 
\end{eqnarray}
with $\eta_{+-} = \eta_{-+} =1$ and $\eta_{ij} = \delta_{ij}$ . 
The non-vanishing components of spin-connections are given by 
\begin{eqnarray} \label{spin}
\omega_{-i} = - \omega_{i-} = \omega^{+i} = - \omega^{i+} = \frac{1}{2} \partial_{i} \mathcal{H} \ du \ . 
\end{eqnarray}
In terms of the frame \eqref{frame}, the background fields \eqref{in X coordinates} take form 
\begin{eqnarray} \label{in terms of frame}
&\hat{\Phi}=& \Phi(u) \ ,  \nonumber\\
&\hat{H}_3 =& f_1(u) \ e^- \wedge e^5 \wedge e^4 -   f_2(u) \ e^- \wedge e^6 \wedge e^7 + f_3(u) \ e^- \wedge e^8 \wedge e^5 \ ,                       \nonumber\\
&\hat{F}_2 =& f_4(u) \ e^- \wedge e^5 \ ,                      \nonumber\\
&\hat{F}_4 =&  f_5(u) \ e^- \wedge e^5 \wedge e^7 \wedge e^6 \ .  
\end{eqnarray}

We will now analyse the spinor conditions in detail. Consider the supersymmetric 
variations of the dilatino and gravitino
\begin{eqnarray}  \label{SUSY} 
&& \delta \hat{\lambda} = \frac{1}{2} \cancel{\partial} \hat{\Phi}  \hat{\epsilon} - \frac{1}{24} 
\cancel{\hat{H}} \sigma_3  \hat{\epsilon} \ + \frac{1}{8} e^{\hat{\Phi}} \Bigg[\frac{3}{2} \cancel{\hat{F}}_2 \left(i \sigma_2\right) + \frac{1}{24} \cancel{\hat{F}}_{4}  \sigma_1 \Bigg] \hat{\epsilon} \ , \nonumber\\
&& \delta \hat{\psi}_{\mu} = D_{\mu}\hat{\epsilon} - \frac{1}{8} \hat{H}_{\mu\nu\rho} \Gamma^{\nu\rho} \sigma_3 \hat{\epsilon} + \frac{1}{8} e^{\hat{\Phi}} \Bigg[\frac{1}{2} \cancel{\hat{F}}_2 \left(i \sigma_2\right) + \frac{1}{24} \cancel{\hat{F}}_{4}  \sigma_1 \Bigg] \Gamma_{\mu}\hat{\epsilon} \ ,
\end{eqnarray}
Here we follow the conventions of \cite{Itsios:2012dc,Itsios:2017nou}. In
particular, we have the covariant derivative 
 $D_{\mu}\hat{\epsilon} = \partial_{\mu}\hat{\epsilon} + \frac{1}{4} \omega_{\mu}^{ab} \Gamma_{ab} \hat{\epsilon}$,  and in addition 
 we use the notation $\cancel{\hat{F}}_n \equiv \hat{F}_{i_{1}...i_{n}} \Gamma^{i_{1}...i_{n}}$. In the above, $\sigma_i$ denote the Pauli matrices.
   The Killing spinor $\hat{\epsilon}$ consists of 
 real Majorana-Weyl spinors $\hat{\epsilon}_{\pm}$,
such that
\begin{eqnarray} \label{killing}
\hat{\epsilon} = \left( \begin{array}{ccc}
\hat{\epsilon}_+ \\
\hat{\epsilon}_-\\
\end{array} \right) \ .
\end{eqnarray}
In type-$IIA$ supergravity, $\hat{\epsilon}$ satisfies $\Gamma_{11} \hat{\epsilon} = - \sigma_3 \hat{\epsilon}$.
%
%
%We will now solve the above spinor conditions. Using the explicit expressions 
%for the spin connections, we find various components of the covariant 
%derivative
%\begin{eqnarray}  \label{co der} 
%D_+ = \partial_+ \ , \ D_-  =  \partial_-  + \frac{1}{4} \partial_{i} \mathcal{H} \ \Gamma_{+i} \ , \  D_i = \partial_i \ ,
%\end{eqnarray}
We also introduce
\begin{eqnarray}  \label{gamma} 
\Gamma^{\pm} = \frac{1}{\sqrt{2}} \Big(\Gamma^9 \pm \Gamma^0\Big) \ . 
\end{eqnarray}

We now proceed to solve the spinor conditions. Substituting the background 
fields in \eqref{SUSY}, and setting the dilatino variation to zero, we 
obtain after some simplification
\begin{eqnarray}  \label{dilatino} 
\Gamma^- \Bigg[\dot{\hat{\Phi}} - \frac{1}{2} \Big(f_1(u) \Gamma^{54} -  f_2(u)  \Gamma^{67} + f_3(u) \Gamma^{85}\Big) \sigma_3 + \frac{e^{\hat{\Phi}}}{4}  \Big(3f_4(u) \Gamma^5  \left(i \sigma_2\right) +  f_5(u) \Gamma^{576} \sigma_1\Big) \Bigg] \hat{\epsilon} = 0 \ . \nonumber\\
\end{eqnarray}
The above condition holds provided $\Gamma^- \hat{\epsilon} = 0$. This 
indicates that, subject to the compatibility with the gravitino variation, 
the pp-wave background \eqref{in X coordinates} preserves 16 supercharges.
We now proceed to varify the spinor condition arising from the variation of
the gravitino. Let us first consider the $\delta\hat{\psi}_+$ variation.
The NS-NS three-form does not have any leg along $e^+$. Together with 
$\Gamma_{+} \hat{\epsilon} = \Gamma^{-} \hat{\epsilon} = 0$, the variation
$\delta\hat{\psi}_+=0$ leads to $\partial_+ \hat{\epsilon} = 0$. Thus, we
find that the Killing spinor $\hat{\epsilon}$ is independent of $v$, 
{\it i.e.} $\hat{\epsilon} = \hat{\epsilon} (u,X^i)$.

Now we focus on the variation $\delta\hat{\psi}_i \ , \ i = 1,..., 8$.
The vanishing of $\delta\hat{\psi}_i$ implies that 
\begin{eqnarray}  \label{psi-i} 
&\partial_i \hat{\epsilon}
  = \Gamma^- \mathcal{R}  \ \hat{\epsilon} \ , 
\end{eqnarray}
where we have introduced the notation 
\begin{eqnarray} \label{R} 
&\mathcal{R} =&   \frac{1}{4} \bigg(f_1(u) \Big(\delta_{i4} \Gamma^5 - \delta_{i5}\Gamma^4\Big) -  f_2(u) \Big(\delta_{i7}\Gamma^6 - \delta_{i6}\Gamma^7\Big) + f_3(u) \Big(\delta_{i5}\Gamma^8 - \delta_{i8}\Gamma^5\Big)\bigg) \sigma_3 \nonumber\\
 &&- \frac{e^{\hat{\Phi}}}{8} \bigg(f_4(u) \Gamma^5 \big(i\sigma_2\big) +  f_5(u) \Gamma^{576} \sigma_1\bigg) \Gamma^i  \ . 
\end{eqnarray}
Now, $\Gamma^-$ anticommutes with $\mathcal{R}$ and $\Gamma^-  \hat{\epsilon} = 0$.
Thus, we have $\partial_i \hat{\epsilon}=0$ leading to
$  \hat{\epsilon}=\chi(u)$ for some $\chi(u)$ satisfying $\Gamma^-  \chi(u) = 0$. Finally, we consider the variation $\delta\hat{\psi}_-=0$.
 Note that, in this case the covariant derivative $D_-$ becomes 
 \begin{eqnarray} \label{D-}
 D_- = \partial_- + \frac{1}{2} F_{ij} X^j \Gamma^{-i} \ .   
\end{eqnarray}
After some simplification, we find that the condition $\delta\hat\psi_-=0$ gives rise to
\begin{eqnarray}  \label{cond} 
 \partial_{u} \chi(u)  - \frac{1}{4} \Big(f_1(u) \Gamma^{54} -  f_2(u)  \Gamma^{67} + f_3(u) \Gamma^{85}\Big) \sigma_3 \chi(u) - \frac{e^{\hat{\Phi}}}{4} \bigg(f_4(u)  \Gamma^5 \big(i\sigma_2\big) +  f_5(u) \Gamma^{576} \sigma_1\bigg)  \chi(u) = 0 \ . \nonumber\\
\end{eqnarray}
This is of the form $\partial_u\chi(u) - {\mathcal M}(u) \chi(u) = 0$, which can 
be integrated to give rise 
$$\chi(u) = e^{\int du {\mathcal M}(u)} \chi_0 \ . $$
This proves that the gravitino condition is compatible with the dilatino 
variation for the above choice of $\chi(u)$, provided $\Gamma^-\chi_0=0$.
Thus, from the above analysis we find that the pp-wave background \eqref{SUSY} 
indeed preserves 16 supercharges.

\section{Gauge theory duals} \label{gauge theory} 

It is well known that the field theory dual to the Klebanov-Tseytlin geometry 
consists of a nonconformal $\mathcal{N}=1$ supersymmetric $SU(N+M)\times SU(N)$ gauge 
theory \cite{Klebanov:2000nc}. It describes the dynamics of $N$ regular and $M$ fractional $D3$ branes
placed near a conifold singularity. The fractional $D3$ branes arise due to 
$D5$ branes wraping the vanishing two-cycle at conifold singularity. The 
nonconformal gauge theory has a nontrivial RG flow. Near UV the supergravity
description is valid and the dual geometry is given by the Klebanov-Tseytlin
background. As the theory is flown to IR it undergoes to a cascade of 
Seiberg dualities there by changing the number of $D3$ branes from $N$ 
to $N-M$ in each step, resulting a singular geometry at the end. However,
for suitably chosen initial condition the conifold 
geometry gets deformed at IR by strong coupling effects there by leading 
to the Klebanov-Strassler background \cite{Klebanov:2000hb}.

For the non-Abelian T-dual geometry of the Klebanov-Tseytlin background the
field theory dual has been considered in \cite{Itsios:2012zv,Itsios:2013wd}.
The existance of domain wall configurations play a key role in understanding
the dual field theories. In the type-$IIB$ theory domain walls can be formed
by wraping $D5$ branes on suitably chosen two cycles of the internal 
manifold. For the Klebanov-Tseytlin background \eqref{KT}-\eqref{KT B2} 
one such two cycle can be constructed upon the identification:
$$\theta_1 = \theta_2, \phi_1 = 2\pi - \phi_2, \psi = \psi_0 \ , $$
with a constant $\psi_0$. This gives rise to the following two cycle 
$\Sigma_2$ for the T-dual background \eqref{rescaled NATD-dual metric}-\eqref{rescaled NATD RR sector}:
\begin{eqnarray} \label{2 cycle}
\Sigma_2 = \Big[\theta_1, \phi_1\Big], \  v_2 = v_3 = \psi = 0 \ .
\end{eqnarray}
Furthermore, it is possible to construct a three cycle $\Sigma_3$ in
the T-dual geometry as 
\begin{eqnarray} \label{3 cycle}
\Sigma_3 = \Big[\theta_1, \phi_1, \psi \Big], \  v_2 = v_3 = r = \text{constant} \ .
\end{eqnarray}

Following \cite{Itsios:2013wd} we will now analyse the construction of 
domain wall in the Klebanov-Tseytlin background and its T-dual. For the
Klebanov-Tseytlin background we will consider the domain wall formed by
a $D5$ brane extended along $\mathbb{R}^{1,2}\in\mathbb{R}^{1,3}$ of the 
$(1,3)$ Minkowski spacetime and wraping the compact directions parametrized
by $\{\theta_2,\phi_2,\psi\}$. The dynamics of the low energy excitations
are captured in terms of the corresponding Born-Infeld action on the world
volume of the $D5$ brane. This gives rise to the corresponding effective 
tension.

The non-Abelian T-duality is performed along an $SU(2)$ isometry parametrized
by the coordinates $\{\theta_2,\phi_2,\psi\}$. Thus the $D5$ brane wraping 
the $SU(2)$ directions gives rise to a $D2$ brane extending along the 
$\mathbb{R}^{1,2}$ of the T-dual geometry. We will place this domain wall
at the origin of the internal manifold:
$$v_2 = v_3 = \theta_1 = \phi_1 = \psi = 0 \ . $$
Once again we can consider the corresponding Born-Infeld action and compute
the effective tension for it. The effective tension of the domain wall in
the Klebanov-Tesytlin background matches with the effective tension of the 
corresponding configuration in the T-dual geometry upto a constant factor
\cite{Itsios:2013wd}. In addition, it has been shown that \cite{Itsios:2013wd} 
the central charge as well as entanglement entropy of both the theories 
match upto an RG independent coefficient. While T-duality maintains the
essential features of the central charge and entanglement entropy, this is 
not the case for the four dimensional gauge coupling. The T-dual geometry
gives rise to a very unusual behaviour for the gauge coupling. It has been
demonstrated that $\nicefrac{1}{g^2}\sim \big(\ln r\big)^{\nicefrac{3}{2}}$, 
unlike the case for a conventional field theory where a logarithmic behaviour
is observed.

The Maxwell and Page charges of the $D$-branes in the theory also play a 
significant role in understanding the field theory dual. For the 
Klebanov-Tseytlin background we consider $D3$ and $D5$ brane charges
\begin{eqnarray} \label{maxwell 2b}
&Q_{\text{Max, D3}} =&  \frac{1}{2k_{10}^2 T_{D3}}  \int_{T^{1,1}} F_5  = \frac{K(r)}{27\pi} \ , \nonumber\\
&Q_{\text{Max, D5}} =&  \frac{1}{2k_{10}^2 T_{D5}}  \int_{\theta_2, \phi_2, \psi} F_3  = \frac{\sqrt{2} P}{9} \ , 
\end{eqnarray}
and 
\begin{eqnarray} \label{page 2b}
&Q_{\text{Page, D3}} =&  \frac{1}{2k_{10}^2 T_{D3}}  \int_{T^{1,1}} F_5 - B_2 \wedge F_3  = \frac{Q(r)}{27\pi} \ , \nonumber\\
&Q_{\text{Page, D5}} =&  \frac{1}{2k_{10}^2 T_{D5}}  \int_{\theta_2, \phi_2, \psi} F_3 - B_2 \wedge F_1  = \frac{\sqrt{2} P}{9} \ , 
\end{eqnarray}
where $Q(r) = K(r) + PT(r)$ and we choose the normalization factor as described in \cite{Macpherson:2014eza}.

For the T-dual background, the Maxwell and Page charges of $D6$ and $D8$ are 
given respectively as 
\begin{eqnarray} \label{maxwell 2a}
&\hat{Q}_{\text{Max, D6}} =&  \frac{1}{\sqrt{2} \pi^2}  \int_{\theta_1, \phi_1} \hat{F}_2  = \frac{K(r)+Q(r)}{27\pi} \ , \nonumber\\
&\hat{Q}_{\text{Max, D8}} =&  \sqrt{2} \int \hat{F}_0   = \frac{\sqrt{2} P}{9} \ , 
\end{eqnarray}
and 
\begin{eqnarray} \label{page 2a}
&\hat{Q}_{\text{Page, D6}} =&  \frac{1}{\sqrt{2} \pi^2}  \int_{\theta_1, \phi_1} \hat{F}_2 - \hat{B}_2 \hat{F}_0 = \frac{2 Q(r)}{27\pi} \ , \nonumber\\
&\hat{Q}_{\text{Page, D8}} =&  \sqrt{2} \int \hat{F}_0   = \frac{\sqrt{2} P}{9} \ .
\end{eqnarray}
The above shows that, after dualization we find $D8$ branes for each of the 
$D5$ branes and twice the number of $D6$ branes for each of the $D3$ branes
in the original background. It has been noticed in \cite{Itsios:2013wd} that
the changes induced in the page charge of $D3$ brane in the Klebanov-Tseytlin
background by a large gauge transformation of the NS-NS two from $B_2$ is 
the same as the changes in the Maxwell charge by a suitable change in the 
radial coordinate. Similar phenomenon is observed in the dual gauge theory,
where the page charge of the $D6$ brane now undergoes a shift under the 
large gauge transformation. This suggests
that the quiver theory corresponding to the T-dual geometry undergoes to a 
cascade of Seiberg dualities much the same way as the gauge theory 
corresponding to the original geometry. Since the change in $D6$ brane charge is twice 
the change in $D3$ charge, the T-dual theory undergoes a Seiberg duality
by a change of $2M$ units of $D6$ brane charge for a change of $M$ units 
of the $D3$ brane charge in the Klebanov-Tseytlin background. 

We will now consider the Maxwell and Page charges for pp-wave background.
Recall that the RR field strengths  for this background in Brinkmann 
coordinates are given as
\begin{eqnarray} \label{Brinkmann RR two}
 &\hat{F}_0=&  0 \ ,  \nonumber\\
 &\hat{F}_2=&   \frac{2^{(- \frac{3}{4})}}{9\sqrt{3} \ \tilde{g}_s} \frac{J}{{\Big(\ln\Big({\frac{r}{\tilde{r}}\Big)\Big)^{\frac{3}{4}}}}} \Bigg[\frac{2}{5} \ln\Big(\frac{r}{\tilde{r}}\Big) + 24 \ln\Big(\frac{r}{r_0}\Big) - \frac{1}{10}\Bigg] du \wedge dz  \ ,       \nonumber\\
 &\hat{F}_4=&   \frac{2^{ \frac{7}{4}}}{3\sqrt{3} \ \tilde{g}_s} \frac{J}{{\Big(\ln\Big({\frac{r}{\tilde{r}}\Big)\Big)^{\frac{5}{4}}}}} \ v_2 \Bigg[-\ln\Big(\frac{r}{\tilde{r}}\Big) \cr && + \frac{1}{2} \ln\Big(\frac{r}{r_0}\Big) \bigg(\frac{1}{15} \ln\Big(\frac{r}{\tilde{r}}\Big) % \nonumber\\
 -\frac{1}{60}\bigg)  \Bigg] du \wedge dz \wedge d\psi \wedge dv_2 \ . 
\end{eqnarray}
The Maxwell and page charge for various brane in type-$IIA$ theory is given by  
\begin{eqnarray} 
&\hat{Q}_{\text{Max, D6}} =&  \frac{1}{\sqrt{2} \pi^2}  \int \hat{F}_2 \ ,   \nonumber\\
&\hat{Q}_{\text{Max, D8}} =&  \sqrt{2} \int \hat{F}_0    \ , 
\end{eqnarray}
and 
\begin{eqnarray}\label{nocascad}
&\hat{Q}_{\text{Page, D6}} =&  \frac{1}{\sqrt{2} \pi^2}  \int \hat{F}_2 - \hat{B}_2 \hat{F}_0  \ , \nonumber\\
&\hat{Q}_{\text{Page, D8}} =&  \sqrt{2} \int \hat{F}_0    \ . 
\end{eqnarray}

Since $\hat F_0$ is zero for our background, the $D8$ charges are all zero.
Moreover, the Maxwell and Page charges for $D6$ branes are both equal. 
The Maxwell and Page charges for $D2$ brenes also vanish. We have
\begin{eqnarray} 
\hat{Q}_{\text{Max, D2}} &=& \frac{1}{2k^2_{10} T_{D2}} \int_{c_6} \hat F_6 \ ,  \cr
\hat{Q}_{\text{Page, D2}} &=& \frac{1}{2k^2_{10} T_{D2}} \int_{c_6}  \bigg[\hat F_6 - \hat B_2 \wedge\hat F_4 + \frac{1}{2}\hat B_2 \wedge\hat B_2 \wedge\hat F_2  \nonumber\\
&&-  \frac{1}{6}\ \hat F_0\hat B_2 \wedge\hat B_2 \wedge\hat B_2\bigg] \ . 
\end{eqnarray}
For fixed $v_2$, both $\hat F_4$ and $\hat F_6$ are zero and $\hat F_0$ as well as 
$\hat B_2\wedge\hat F_2$ vanish for the pp-wave background. From \eqref{nocascad},
we find that there is no longer any cascading due to large gauge transformation 
of $\hat B_2$. This indicates that the quiver theory dual to the pp-wave geometry
correspond to the end point of the cascade.

\section{Conclusion}  

In this paper we have considered Penrose limits for the Klebaov-Tseytlin 
geometry and its non-Abelian T-dual around a suitable $SU(2)$ isometry.
We have scrutinized various null geodesics in these geometries. A direct investigation 
of the Penrose limits for the Klebanov-Tseytlin geometry gives rise to 
singular geometries for most of the null geodesics. We found one smooth 
geometry with a nonvanishing scalar curvature. However upon taking the 
non-Abelian T-duality results a pp-wave solution around a suitably chosen
null geodesic. The holographic dual of the T-dual background exhibits 
a cascade of Seiberg dualities under large gauge transformation of the
NS-NS two form there by reducing the number of $D6$-branes in each step.
However, an analysis of the Maxwell and Page chrages shows the absence
of similar phenomenon for the pp-wave background. Thus, the holographic 
dual in this case appears to be the end point of the cascade of quivers
corresponding to the T-dual geometry. The gauge coupling analysis shows
that the quiver in the T-dual case is a non-conventional field theory.
Further investigation is required to precisely identify the quiver and 
also the corresponding BMN sector and to establish a map between holographic
quantities and field theory observables. It would also be interesting to 
explore the possibility of obtaining pp-wave geometries for the non-Abelian 
T-dual of Klebanov-Strassler background as well as backgrounds with AdS$_3$ 
factors. Dualization of the Baryonic branch of the Klebanov-Strassler geometry
has already been carried out \cite{Gaillard:2013vsa}.
 We hope to address some of these issues in future.

\appendix

\section{Penrose limits in Klebanov-Tseytlin background}

In this appendix we will examine Penrose limits for various null geodesics 
in the Klebanov-Tseytlin geometry. Recall the metric of Klebanov-Tseytlin background 
\begin{eqnarray} 
ds^2 =  H(r)^{-\frac{1}{2}} \ \eta_{\mu\nu} dx^\mu dx^\nu + H(r)^{\frac{1}{2}} \Big(dr^2 + r^2 ds_{T^{1,1}}^2\Big) \ ,
\end{eqnarray}
with the warp factor
\begin{eqnarray} 
H(r) =  \frac{1}{r^4} \Bigg[R^4 + 2L^4 \Bigg(\ln\bigg(\frac{r}{r_0}\bigg) + \frac{1}{4}\Bigg)\Bigg] = \frac{2L^4}{r^4} \ln\bigg(\frac{r}{\tilde{r}}\bigg)  \ . 
\end{eqnarray}
The $T^{1,1}$ metric is given by 
\begin{eqnarray} 
ds^2_{T^{1,1}}= \lambda_{1}^2\ d\Omega^2_{2} \big(\theta_{1},\phi_{1}\big) + \lambda_{2}^2\ d\Omega^2_{2} \big(\theta_{2},\phi_{2}\big)
+ \lambda^2\big(d\psi + \cos\theta_{1} d\phi_{1} + \cos\theta_{2} d\phi_{2}\big)^2 \ .
\end{eqnarray}

The parameters $\lambda,\lambda_1,\lambda_2$ in the $T^{1,1}$ metric have the numerical values
  $\lambda_{1}^2=\lambda_{2}^2=\frac{1}{6}$ and $\lambda^2=\frac{1}{9}$ .

We will now rescale the Minkowski coordinates ($x^\mu$) as $x^\mu \rightarrow L^2 x^\mu$.
The metric then becomes 
\begin{eqnarray} \label{rescaled metric}
ds^2 = L^2 \Bigg[\frac{1}{\sqrt{2}} \frac{r^2}{\sqrt{\ln\Big(\frac{r}{\tilde{r}}\Big)}}  \ \eta_{\mu\nu} dx^\mu dx^\nu + \frac{\sqrt{2}}{r^2} \sqrt{\ln\Big(\frac{r}{\tilde{r}}\Big)} \ dr^2  + \sqrt{2} \ \sqrt{\ln\Big(\frac{r}{\tilde{r}}\Big)} \ ds^2_{T^{1,1}}\Bigg] \ .  
\end{eqnarray}
The above metric has $U(1)$ isometries along $\phi_1 , \phi_2$ and $\psi$ directions.
 We will examine the Penrose limits by considering the motion along these isometry directions. Let us first consider motion along $\psi$-direction.
The ${\psi\psi}$-component of the metric is given by 
\begin{eqnarray} 
g_{\psi\psi} = \lambda^2 L^2 \sqrt{2}  \ \sqrt{\ln\Big(\frac{r}{\tilde{r}}\Big)}  \ . 
\end{eqnarray}
Now by imposing the null geodesic condition we get
\begin{eqnarray} 
\lambda^2 L^2  \frac{1}{\sqrt{2} r \sqrt{\ln\Big(\frac{r}{\tilde{r}}\Big)}}  = 0 \ ,
\end{eqnarray}
which does not admit any smooth solution.

Now we shall consider motion along the $\phi_1$-direction.
The relevant metric component is
\begin{eqnarray} 
g_{\phi_1\phi_1} =  L^2 \sqrt{2}  \ \sqrt{\ln\Big(\frac{r}{\tilde{r}}\Big)} \bigg[\lambda_1^2 \sin^2\theta_1 + \lambda^2 \cos^2\theta_1\bigg] \ , 
\end{eqnarray}
The geodesic condition for $\mu = r$ does not give any solution. For $\mu = \theta_1$ the condition leads to the solution $r = \tilde{r}$ and 
$\theta_1 = \{0, \frac{\pi}{2}, \pi\}$.
However, $r = \tilde{r}$ is a singular point as the metric component $g_{\phi_1\phi_1}$ vanishes for this value of $r$.
Hence we shall consider Penrose limit around $\theta_1 = \{0, \frac{\pi}{2}, \pi\} , \theta_2 = \phi_2 = \psi =0$ while the $r$-coordinate fixed.

Consider the following expansion around the geodesic $\theta_1 = \theta_2 = \phi_2 = \psi =0$ and $r =$ constant:
\begin{eqnarray} 
&&x_i  = \frac{y_i}{L} \ ; \ i =1, 2, 3, \  r = c + \frac{w}{L} \ , \ \theta_1 =  \frac{z}{L} \ ,  \ \theta_2 =  \frac{x}{L} \ , \nonumber\\ 
&&t = ax^+ \ , \ \phi_1 = bx^+ + \frac{x^-}{L^2} \ , \ \phi_2 \rightarrow \frac{\phi_2}{L} \ , \ \psi \rightarrow \frac{\psi}{L} \ ,  
\end{eqnarray}
where $a, b\ \&\ c$ are some nonvanishing parameters. 
The null geodesic condition gives 
$$a^2 = \frac{2 \lambda^2 b^2}{c^2} \ln\Big(\frac{c}{\tilde{r}}\Big) \ . $$
For the above expansion, the leading terms of the T-dual metric in the limit $L\rightarrow \infty$ are given by 
\begin{eqnarray} 
&ds^2=&  2\sqrt{2} \lambda^2 b  \ \sqrt{\ln\Big({\frac{c}{\tilde{r}}\Big)}} \ dx^+ dx^- + \frac{1}{\sqrt{2}} \ \frac{c^2}{\sqrt{\ln\Big({\frac{c}{\tilde{r}}\Big)}}} \Big(dy_1^2 + dy_2^2 + dy_3^2\Big) +\frac{\sqrt{2}}{c^2}\  \sqrt{\ln\Big({\frac{c}{\tilde{r}}\Big)}}  \ dw^2 \nonumber\\
&& + \ \lambda_1^2 \sqrt{2} \ \sqrt{\ln\Big({\frac{c}{\tilde{r}}\Big)}} \ dz^2 +  \lambda_2^2 \sqrt{2} \ \sqrt{\ln\Big({\frac{c}{\tilde{r}}\Big)}} \ dx^2 + \lambda^2 \sqrt{2} \ \sqrt{\ln\Big({\frac{c}{\tilde{r}}\Big)}} \bigg(\Big(d\psi + d\phi_2\Big)^2 +   \nonumber\\
&&  \frac{wb}{c \ln\Big(\frac{c}{\tilde{r}}\Big)} \ dx^+ d\psi+ \ 2b dx^+ d\phi_2 \bigg)+ \  \Bigg[ \frac{1}{\sqrt{2}} \ \frac{-w^2a^2}{\sqrt{\ln\Big({\frac{c}{\tilde{r}}\Big)}}} \bigg(\frac{1}{4\ln\Big(\frac{c}{\tilde{r}}\Big)} +1 \bigg) \ + \sqrt{2} \sqrt{\ln\Big({\frac{c}{\tilde{r}}\Big)}}  \ b^2 z^2 \Big(\lambda_1^2 - \lambda^2\Big)   \nonumber\\ 
&& - \ \frac{w^2b^2}{4c^2\ln\Big(\frac{c}{\tilde{r}}\Big)}\Bigg](dx^+)^2  +  L \Bigg[\frac{1}{\sqrt{2}} \ \frac{cwa^2}{\sqrt{\ln\Big({\frac{c}{\tilde{r}}\Big)}}}  \bigg(\frac{1}{2 \ln\Big(\frac{c}{\tilde{r}}\Big)} - 2\bigg) (dx^+)^2 + 2\sqrt{2} \lambda^2 b  \ \sqrt{\ln\Big({\frac{c}{\tilde{r}}\Big)}} \ dx^+ d\psi\Bigg] \ . \nonumber\\
\end{eqnarray}
We can see that the divergent term of order $\mathcal{O}(L)$ can't be removed for any
choice of the parameters $a,b,c$.

Now we consider the following expansion around the geodesic $\theta_1 = \frac{\pi}{2}$ and $\theta_2 = \phi_2 = \psi =0$ and $r =$ constant:
\begin{eqnarray} 
&&x_i  = \frac{y_i}{L} \ ; \ i =1, 2, 3, \  r = c + \frac{w}{L} \ , \ \theta_1 = \frac{\pi}{2} +  \frac{z}{L} \ ,  \ \theta_2 =  \frac{x}{L} \ , \ t = ax^+ \ , \nonumber\\
&&\phi_1 = bx^+ + \frac{x^-}{L^2} \ , \ \phi_2 \rightarrow \frac{\phi_2}{L} \ , \ \psi \rightarrow \frac{\psi}{L} \ .
 \end{eqnarray}
The null geodesic condition gives 
$$a^2 = \frac{2 \lambda_1^2 b^2}{c^2} \ln\Big(\frac{c}{\tilde{r}}\Big) \ . $$
The leading terms of the T-dual metric in the limit $L\rightarrow \infty$ are given by 
\begin{eqnarray} 
&ds^2=&  2\sqrt{2} \lambda_1^2 b  \ \sqrt{\ln\Big({\frac{c}{\tilde{r}}\Big)}} \ dx^+ dx^- + \frac{1}{\sqrt{2}} \ \frac{c^2}{\sqrt{\ln\Big({\frac{c}{\tilde{r}}\Big)}}} \Big(dy_1^2 + dy_2^2 + dy_3^2\Big) +\frac{\sqrt{2}}{c^2}\  \sqrt{\ln\Big({\frac{c}{\tilde{r}}\Big)}}  \ dw^2 \nonumber\\
&& + \ \lambda_1^2 \sqrt{2} \ \sqrt{\ln\Big({\frac{c}{\tilde{r}}\Big)}} \ dz^2 +  \lambda_2^2 \sqrt{2} \ \sqrt{\ln\Big({\frac{c}{\tilde{r}}\Big)}} \ dx^2 + \lambda^2 \sqrt{2} \ \sqrt{\ln\Big({\frac{c}{\tilde{r}}\Big)}} \bigg(\Big(d\psi + d\phi_2\Big)^2   \nonumber\\
&&  - 2bzdx^+ d\psi -  2bz dx^+ d\phi_2 \bigg)+ \  \Bigg[ \frac{1}{\sqrt{2}} \ \frac{-w^2a^2}{\sqrt{\ln\Big({\frac{c}{\tilde{r}}\Big)}}} \bigg(\frac{1}{4\ln\Big(\frac{c}{\tilde{r}}\Big)} +1 \bigg) \ + \sqrt{2} \sqrt{\ln\Big({\frac{c}{\tilde{r}}\Big)}}  \ b^2 z^2 \Big(\lambda^2 - \lambda_1^2\Big)   \nonumber\\ 
&& - \frac{w^2b^2}{4c^2\ln\Big(\frac{c}{\tilde{r}}\Big)}\Bigg] (dx^+)^2 + L \Bigg[\frac{1}{\sqrt{2}} \ \frac{cwa^2}{\sqrt{\ln\Big({\frac{c}{\tilde{r}}\Big)}}}  \bigg(\frac{1}{2 \ln\Big(\frac{c}{\tilde{r}}\Big)} - 2\bigg)  + \frac{\lambda_1^2}{c\sqrt{2}} \frac{wb^2}{\sqrt{\ln\Big({\frac{c}{\tilde{r}}\Big)}}} \Bigg] (dx^+)^2 \ . 
\end{eqnarray}
Once again, the metic is divergent in the limit $L\rightarrow \infty$ due to presence of $\mathcal{O}(L)$ term.

We will now consider a null geodesic which carries angular momentum.
To obtain such a geodesic, we consider motion along $r$ and $\phi_1$ directions and concentrate in a small neighbourhood of  $\theta_1 = \theta_2 = \phi_2 = \psi = 0$. The Lagrangian for a massless particle moving along this geodesic is 
\begin{eqnarray}
\mathcal{L} = \frac{1}{2} g_{\mu\nu} \dot{X}^\mu \dot{X}^\nu  \ .
\end{eqnarray}
Here dots denote derivative with respect to the affine parameter $u$.
Substituting the explicit expression for the background metric \eqref{rescaled metric} in the above Lagrangian we find
\begin{eqnarray}
\mathcal{L} = \frac{L^2}{2} \Bigg(-  \frac{1}{\sqrt{2}} \frac{r^2}{\sqrt{\ln\Big({\frac{r}{\tilde{r}}\Big)}}} \ \dot{t}^2 + \frac{\sqrt{2}}{r^2} \sqrt{\ln\Big({\frac{r}{\tilde{r}}\Big)}} \ \dot{r}^2 +  \lambda^2 \sqrt{2} \ \sqrt{\ln\Big({\frac{r}{\tilde{r}}\Big)}} \ \dot{\phi}_1^2\Bigg) \ .
\end{eqnarray}
We notice that the above Lagrangian does not depend on $t$ and $\phi_1$
explicitly. Hence the momenta conjugate to the generalized coordinates 
$t$ and $\phi_1$ are conserved. Denoting these quantities by $E$ and $J$
(upto a factor of $-L^2$), we find
\begin{eqnarray} 
\frac{\partial \mathcal{L}}{\partial\dot{t}} &=& -  \frac{L^2}{\sqrt{2}} \frac{r^2}{\sqrt{\ln\Big({\frac{r}{\tilde{r}}\Big)}}}   \ \dot{t} = -E L^2  \ , \cr
\frac{\partial \mathcal{L}}{\partial\dot{\phi}_1} &=&  \lambda^2 L^2 \sqrt{2}  \ \sqrt{\ln\Big({\frac{r}{\tilde{r}}\Big)}} \ \dot{\phi}_1 =  - J L^2  \ .
\end{eqnarray}
The condition that the geodesic becomes null gives rise to 
\begin{eqnarray}
\dot{r}^2 +  \frac{J^2 r^2}{2 \lambda^2 \ln\Big({\frac{r}{\tilde{r}}\Big)}}  = E^2  \ .
\end{eqnarray}

To obtain the Penrose limit, we redefine the coordinates as
\begin{eqnarray} 
x_i  = \frac{y_i}{L} \ ; \ i =1, 2, 3, \ \theta_1 =  \frac{z}{L} \ , \ \theta_2 =  \frac{x}{L} \ ,\ \phi_2 \rightarrow \frac{\phi_2}{L} \ , \ \psi \rightarrow \frac{\psi}{L} \ ,
\end{eqnarray}
and consider the following expansion in the limit $L \rightarrow \infty$:
\begin{eqnarray} 
dt = c_1 du , \ dr = c_2 du + c_3 \frac{dw}{L}  \ , \ d\phi_1 = c_4 du + c_5 \frac{dw}{L} + c_6 \frac{dv}{L^2}  \  .
\end{eqnarray}

By requiring the geodesic to be null determines the coefficients $c_1, c_2$ and $c_4$ as
\begin{eqnarray} 
c_1 = \frac{E \sqrt{2}}{r^2} \ \sqrt{\ln\Big({\frac{r}{\tilde{r}}\Big)}}   \ , \ c_2 = \Bigg[E^2 -  \frac{J^2r^2}{2 \lambda^2 \ln\Big({\frac{r}{\tilde{r}}\Big)}} \Bigg]^{\frac{1}{2}}  , \ c_4 = -  \frac{J}{\lambda^2 \sqrt{2} \sqrt{\ln\Big({\frac{r}{\tilde{r}}\Big)}}}   \  .
\end{eqnarray}
The metric then becomes
\begin{eqnarray} 
&ds^2=&  2\sqrt{2} \lambda^2 \ \sqrt{\ln\Big({\frac{r}{\tilde{r}}\Big)}} \ c_4 c_6 \ du dv + \frac{1}{\sqrt{2}} \ \frac{r^2}{\sqrt{\ln\Big({\frac{r}{\tilde{r}}\Big)}}} \Big(dy_1^2 + dy_2^2 + dy_3^2\Big) \nonumber\\
&& + \ \sqrt{2}\  \sqrt{\ln\Big({\frac{r}{\tilde{r}}\Big)}} \ \bigg(\frac{c_3^2}{r^2} +\lambda^2 c_5^2 \bigg) dw^2  + \lambda_1^2 \sqrt{2} \ \sqrt{\ln\Big({\frac{r}{\tilde{r}}\Big)}} \ dz^2 +  \lambda_2^2 \sqrt{2} \ \sqrt{\ln\Big({\frac{r}{\tilde{r}}\Big)}} \ dx^2  \nonumber\\
&& + \ \lambda^2 \sqrt{2} \ \sqrt{\ln\Big({\frac{r}{\tilde{r}}\Big)}} \bigg(\Big(d\psi + d\phi_2\Big)^2 + 2c_3 dw \Big(d\psi + d\phi_2\Big)\bigg) +  \sqrt{2} \ \sqrt{\ln\Big({\frac{r}{\tilde{r}}\Big)}}  \ c_4^2 z^2 \Big(\lambda_1^2 - \lambda^2\Big) du^2  \nonumber\\
&& + \  L \Bigg[\frac{2\sqrt{2}}{r^2} \ \sqrt{\ln\Big({\frac{r}{\tilde{r}}\Big)}} \ \Big(c_2 c_3 + \lambda^2 r^2 c_4 c_5 \Big) du dw +  2 c_4 \Big(d\psi + d\phi_2\Big) du\Bigg] \ .  
\end{eqnarray}
We note that the divergent term  of order $\mathcal{O}(L)$ can't be removed. Thus the 
Penrose limit around this geodesic does not give a smooth geometry.

Finally, we will consider the expansion around $\theta_1 = \frac{\pi}{2}$ and concentrate in a small neighbourhood of  $\theta_2 = \phi_2 = \psi = 0$ .
The Lagrangian for a massless particle then gives rise to 
\begin{eqnarray}
\mathcal{L} = \frac{L^2}{2} \Bigg(-  \frac{1}{\sqrt{2}} \frac{r^2}{\sqrt{\ln\Big({\frac{r}{\tilde{r}}\Big)}}} \ \dot{t}^2 + \frac{\sqrt{2}}{r^2} \sqrt{\ln\Big({\frac{r}{\tilde{r}}\Big)}} \ \dot{r}^2 +  \lambda_1^2 \sqrt{2} \ \sqrt{\ln\Big({\frac{r}{\tilde{r}}\Big)}} \ \dot{\phi}_1^2\Bigg) \ .
\end{eqnarray}
The conserved quantities are 
\begin{eqnarray} 
\frac{\partial \mathcal{L}}{\partial\dot{t}} &=& -  \frac{L^2}{\sqrt{2}} \frac{r^2}{\sqrt{\ln\Big({\frac{r}{\tilde{r}}\Big)}}}   \ \dot{t} = -E L^2  \ , \cr
\frac{\partial \mathcal{L}}{\partial\dot{\phi}_1} &=&  \lambda_1^2 L^2 \sqrt{2}  \ \sqrt{\ln\Big({\frac{r}{\tilde{r}}\Big)}} \ \dot{\phi}_1 =  - J L^2  \ ,
\end{eqnarray}
and the null geodesic condition becomes
\begin{eqnarray}
\dot{r}^2 +  \frac{J^2 r^2}{2 \lambda_1^2 \ln\Big({\frac{r}{\tilde{r}}\Big)}}  = E^2  \ .
\end{eqnarray}

To obtain the Penrose limit, we redefine the coordinates 
\begin{eqnarray} 
x_i  = \frac{y_i}{L} \ ; \ i =1, 2, 3, \ \theta_1 =  \frac{\pi}{2} + \frac{z}{L} \ , \ \theta_2 =  \frac{x}{L} \ ,\ \phi_2 \rightarrow \frac{\phi_2}{L} \ , \ \psi \rightarrow \frac{\psi}{L} \ ,
\end{eqnarray}
Then consider the following expansion in the limit $L \rightarrow \infty$:
\begin{eqnarray} 
dt = c_1 du , \ dr = c_2 du + c_3 \frac{dw}{L}  \ , \ d\phi_1 = c_4 du + c_5 \frac{dw}{L} + c_6 \frac{dv}{L^2}  \  .
\end{eqnarray}
The null geodesic condition determines the coefficients $c_1, c_2$ and $c_4$ as
\begin{eqnarray} 
c_1 = \frac{E \sqrt{2}}{r^2} \ \sqrt{\ln\Big({\frac{r}{\tilde{r}}\Big)}}   \ , \ c_2 = \Bigg[E^2 -  \frac{J^2r^2}{2 \lambda^2 \ln\Big({\frac{r}{\tilde{r}}\Big)}} \Bigg]^{\frac{1}{2}}  , \ c_4 = -  \frac{J}{\lambda_1^2 \sqrt{2} \sqrt{\ln\Big({\frac{r}{\tilde{r}}\Big)}}}   \  .
\end{eqnarray}
The background metric then takes the form
\begin{eqnarray} 
&ds^2=&  2\sqrt{2} \lambda_1^2 \ \sqrt{\ln\Big({\frac{r}{\tilde{r}}\Big)}} \ c_4 c_6 \ du dv + \frac{1}{\sqrt{2}} \ \frac{r^2}{\sqrt{\ln\Big({\frac{r}{\tilde{r}}\Big)}}} \Big(dy_1^2 + dy_2^2 + dy_3^2\Big)   \nonumber\\
&& + \sqrt{2}\  \sqrt{\ln\Big({\frac{r}{\tilde{r}}\Big)}} \ \bigg(\frac{c_3^2}{r^2} +\lambda_1^2 c_5^2 \bigg) dw^2  + \lambda_1^2 \sqrt{2} \ \sqrt{\ln\Big({\frac{r}{\tilde{r}}\Big)}} \ dz^2 +  \lambda_2^2 \sqrt{2} \ \sqrt{\ln\Big({\frac{r}{\tilde{r}}\Big)}} \ dx^2  \nonumber\\
&& + \lambda^2 \sqrt{2} \ \sqrt{\ln\Big({\frac{r}{\tilde{r}}\Big)}} \bigg(\Big(d\psi + d\phi_2\Big)^2 - 2c_4 z \ du \Big(d\psi + d\phi_2\Big)\bigg) +   \sqrt{2} \ \sqrt{\ln\Big({\frac{r}{\tilde{r}}\Big)}}  \ c_4^2 z^2 \Big(\lambda^2 - \lambda_1^2\Big) du^2 \ .  \nonumber\\ 
\end{eqnarray}
along with a order $\mathcal{O}(L)$ divergent term which can be removed upon 
requiring 
$$c_2 c_3 + \lambda_1^2 r^2 c_4 c_5 = 0 \ , $$
there by giving rise to a smooth geometry. However, a straightforward calculation
gives rise to a nonvanishing scalar curvature for this geometry. Hence, this 
does not correspond to a pp-wave.

\section{Einstein's Equations}

In this appendix, we will analyse the Einstein's equations for our pp-wave
background. The Einstein's equation for type-$IIA$ supergravity is given by 
\begin{eqnarray} \label{Einstein eq natd1}
\hat{R}_{\mu\nu} + 2D_{\mu}D_{\nu}\hat{\Phi} = \frac{1}{4} \hat{H}_{\mu\nu}^2 + e^{2\hat{\Phi}} \Bigg[\frac{1}{2} (\hat{F}_2^2)_{\mu\nu} + \frac{1}{12} (\hat{F}_4^2)_{\mu\nu} - \frac{1}{4} g_{\mu\nu} \Big(\hat{F}_{0}^2 + \frac{1}{2} \hat{F}_{2}^2 + \frac{1}{4!}\hat{F}_{4}^2 \Big)\Bigg]  \ .  
\end{eqnarray}
Here we use the conventions of \cite{Itsios:2012dc}. In particular, we have
$\hat{H}_{\mu\nu}^2 = \hat{H}_{\mu\alpha\beta}\hat H_{\nu\rho\sigma}
g^{\alpha\rho}g^{\beta\sigma}$ and similar expressions for 
$(\hat{F}_2^2)_{\mu\nu}$ and $(\hat{F}_4^2)_{\mu\nu}$.
The equation of motion for the dilation is given by
\begin{eqnarray} \label{R eq natd1}
\hat{R} + 4D^2\hat{\Phi} - 4(\partial\hat{\Phi})^2 - \frac{1}{12}\hat{H}^2=0 \ .
\end{eqnarray}
We will first focus on the dilation equation. For the pp-wave background 
$\hat R = 0$. Now, consider evaluating $D^2\hat\Phi$. Note that
\begin{eqnarray} 
&D^2 \hat{\Phi} =& g^{\mu\nu} D_{\mu} D_{\nu} \hat{\Phi} = g^{uv} D_{u} D_{v} \hat{\Phi} + g^{vu} D_{v} D_{u} \hat{\Phi} + g^{vv} D_{v} D_{v} \hat{\Phi} + g^{ij} D_{i} D_{j} \hat{\Phi} \ , 
\end{eqnarray}
for $i,j\neq\{u,v\}$. Consider the covariant derivatives of the form 
$D_{\mu} D_{\nu} \hat{\Phi}$ appearing in the above equation. Since,
$\partial_v\hat\Phi = 0 = \partial_i\hat\Phi$, we find
\begin{eqnarray}
&& D_{u} D_{v} \hat{\Phi} = \partial_u \partial_v \hat{\Phi} - \Gamma_{uv}^{\lambda} \partial_\lambda \hat{\Phi} = - \Gamma_{uv}^{u} \partial_u \hat{\Phi} \ , \nonumber\\
&& D_{v} D_{u} \hat{\Phi} = \partial_v \partial_u \hat{\Phi} - \Gamma_{vu}^{\lambda} \partial_\lambda \hat{\Phi} = - \Gamma_{vu}^{u} \partial_u \hat{\Phi} \  , \nonumber\\
&& D_{v} D_{v} \hat{\Phi} = \partial_v \partial_v \hat{\Phi} - \Gamma_{vv}^{\lambda} \partial_\lambda \hat{\Phi} = - \Gamma_{vv}^{u} \partial_u \hat{\Phi} \  , \nonumber\\
&& D_{i} D_{j} \hat{\Phi} = \partial_i \partial_j \hat{\Phi} - \Gamma_{ij}^{\lambda} \partial_\lambda \hat{\Phi} = - \Gamma_{ij}^{u} \partial_u \hat{\Phi} \ .
\end{eqnarray}
It is straightforward to evaluate the Christoffel symbols. We find 
\begin{eqnarray} 
&&\Gamma_{uv}^u = \frac{1}{2} g^{uv} \Big(\partial_v g_{vu} + \partial_u g_{vv} - \partial_v g_{uv}\Big) = 0  \ , \nonumber\\
&&\Gamma_{vv}^u = \frac{1}{2} g^{uv} \Big(\partial_v g_{vv} + \partial_v g_{vv} - \partial_v g_{uv}\Big) = 0 \ , \nonumber\\
&&\Gamma_{ij}^u = \frac{1}{2} g^{uv} \Big(\partial_j g_{vi} + \partial_i g_{vj} - \partial_v g_{ij}\Big) = 0 \ . 
\end{eqnarray}
Thus, we have $D^2\hat\Phi=0$. Similarly, we can show that 
$\Big(\partial \hat{\Phi}\Big)^2$ also vanishes identically:
\begin{eqnarray} 
\Big(\partial \hat{\Phi}\Big)^2 =& g^{\mu\nu} \partial_{\mu} \hat{\Phi} \partial_{\nu} \hat{\Phi} = 2 g^{uv} \partial_{u}  \hat{\Phi} \partial_{v} \hat{\Phi} + g^{vv} \partial_{v}  \hat{\Phi}  \partial_{v} \hat{\Phi} + g^{ij} \partial_{i}  \hat{\Phi} \partial_{j} \hat{\Phi} = 0 \ . 
\end{eqnarray}
Further, from the expression for $\hat H_3$, we find that $\hat H_3^2=0$. 
This shows that the dilaton equation is satisfied identically.

We will now consider the Einstein's equations. Clearly, from \eqref{Brinkmann RR} 
we have $\hat{F}_0^2 = \hat{F}_2^2 = \hat{F}_4^2 = 0$. Further, a straightforward 
calculation shows that the only the $uu$-components of $ \hat{H}_{\mu\nu}^2 \ ,    (\hat{F}_2)^2 _{\mu\nu}$ , $(\hat{F}_4)^2 _{\mu\nu}$ together with $D_{u} D_{u} \hat{\Phi}$ are non-vanishing. Likewise, we know from  \cite{blau.m} that in
Brinkmann coordinates, the $uu$-component $\hat R_{uu}$ is the only nonvanishing
component of the Ricci tensor. Thus, for our background the Einstein's equation
reduces to 
\begin{eqnarray} \label{Einstein uu1}
\hat{R}_{uu} + 2D_{u}D_{u}\hat{\Phi} = \frac{1}{4} \hat{H}_{uu}^2 + e^{2\hat{\Phi}} \Bigg[\frac{1}{2} (\hat{F}_2^2)_{uu} + \frac{1}{12} (\hat{F}_4^2)_{uu} \Bigg]  \ .
\end{eqnarray}
In the following we evaluate each of the terms of the above equation. The 
nonzero component of the Ricci tensor is
\begin{eqnarray} \label{Ruu}
\hat{R}_{uu} =  - \Bigg[F_{y_{1}} + F_{y_{2}} + F_{y_{3}} + F_w + \Big(F_z -  \frac{3\sqrt{2}}{\sqrt{\ln\Big({\frac{r}{\tilde{r}}\Big)}}} J^2\Big) + F_{v_{2}} + F_{v_{3}}\Bigg] \ , 
\end{eqnarray}
where the expressions for the functions $F_{i}$ are given in \eqref{Fi definitions }. Similarly, the remaining terms in the equation are evaluated to be
\begin{eqnarray} \label{uu}
&e^{-2\hat{\Phi}}=&  \frac{4 \sqrt{2} }{81 \ \tilde{g}_s^2}  \ \sqrt{\ln\Big({\frac{r}{\tilde{r}}\Big)}}  \ \Bigg[\ln\Big({\frac{r}{\tilde{r}}\Big)} + 2 \Big(\ln\Big(\frac{r}{r_0}\Big)\Big)^2 \Bigg] \ , \nonumber\\
&D_{u} D_{u} \hat{\Phi}=& \frac{1}{8r^2 \Big(\ln\Big({\frac{r}{\tilde{r}}\Big)}\Big)^3 \Bigg[\ln\Big({\frac{r}{\tilde{r}}\Big)} + 2 \Big(\ln\Big(\frac{r}{r_0}\Big)\Big)^2 \Bigg]^2} \ f(r) \ , \nonumber\\
&\hat{H}_{uu}^2=&  2 \Bigg[ \frac{\sqrt{2}}{3} \bigg(E^2 -  \frac{3r^2}{\ln\Big({\frac{r}{\tilde{r}}\Big)}} J^2 \bigg)  \bigg(c_5^{\prime} \ln({\frac{r}{r_0})} + \frac{c_5}{r}\bigg)^2 \bigg(\frac{c_3^2 \sqrt{2}}{r^2} + \frac{c_5^2}{3\sqrt{2}} \bigg)^ {-1}  \frac{1}{\ln\Big(\frac{r}{\tilde{r}}\Big)}        \nonumber\\
&& + 2 \bigg(E^2 -  \frac{3r^2}{\ln\Big({\frac{r}{\tilde{r}}\Big)}} J^2 \bigg)  \Bigg(\frac{\ln\Big(\frac{r}{\tilde{r}}\Big) -\ln\Big(\frac{r}{r_0}\Big) - 2\Big(\ln\Big(\frac{r}{r_0}\Big)\Big)^2  }{r \Big(\ln\Big({\frac{r}{\tilde{r}}\Big)} + 2 \Big(\ln\Big(\frac{r}{r_0}\Big)\Big)^2\Big)}\Bigg)^2 \frac{1}{\ln\Big(\frac{r}{\tilde{r}}\Big)} + \frac{12J^2}{\ln\Big(\frac{r}{\tilde{r}}\Big)}\Bigg] \ ,    \nonumber\\
 &(\hat{F}_2)^2 _{uu}=& \frac{2^{(- \frac{3}{2})}}{243 \ \tilde{g}_s^2} \frac{J^2}{{\Big(\ln\Big({\frac{r}{\tilde{r}}\Big)\Big)^{\frac{3}{2}}}}} \Bigg[\frac{2}{5} \ln\Big(\frac{r}{\tilde{r}}\Big) + 24 \ln\Big(\frac{r}{r_0}\Big) - \frac{1}{10}\Bigg]^2                 \ ,              \nonumber\\
&(\hat{F}_4)^2 _{uu}=& \frac{2^{ \frac{9}{2}}}{9 \ \tilde{g}_s^2} \frac{J^2}{{\Big(\ln\Big({\frac{r}{\tilde{r}}\Big)\Big)^{\frac{5}{2}}}}} \  \Bigg[- \ln\Big(\frac{r}{\tilde{r}}\Big) + \frac{1}{2} \ln\Big(\frac{r}{r_0}\Big) \bigg(\frac{1}{15} \ln\Big(\frac{r}{\tilde{r}}\Big) -\frac{1}{60}\bigg)  \Bigg]^2   \ .  
\end{eqnarray}
where the function $f(r)$ has the complicated expression:
\begin{eqnarray} \label{fr}
&f(r)=&  -36J^2r^2 \Big(\ln\Big(\frac{r}{r_0}\Big)\Big)^4 + 2E^2  \Big(\ln\Big(\frac{r}{\tilde{r}}\Big)\Big)^4 \Big(-5 + 8 \ln\Big(\frac{r}{r_0}\Big)\Big)   \nonumber\\
&&+ 8 \ln\Big(\frac{r}{\tilde{r}}\Big) \Big(\ln\Big(\frac{r}{r_0}\Big)\Big)^2  \Big(-6J^2r^2 - 6J^2r^2  \ln\Big(\frac{r}{r_0}\Big) + E^2 \Big(\ln\Big(\frac{r}{r_0}\Big)\Big)^2\Big)                \nonumber\\ 
&&+ \Big(\ln\Big(\frac{r}{\tilde{r}}\Big)\Big)^3 \Big(6\big(E^2 + 8J^2r^2\big) + 32E^2 \ln\Big(\frac{r}{r_0}\Big) + 48E^2  \Big(\ln\Big(\frac{r}{r_0}\Big)\Big)^2 + 32E^2    \Big(\ln\Big(\frac{r}{r_0}\Big)\Big)^3\Big) \nonumber\\
&&+ \Big(\ln\Big(\frac{r}{\tilde{r}}\Big)\Big)^2 \Big(-27J^2r^2 -120 J^2r^2 \ln\Big(\frac{r}{r_0}\Big) + 8\Big(E^2 - 12J^2r^2\Big) \Big(\ln\Big(\frac{r}{r_0}\Big)\Big)^2 + 8E^2 \Big(\ln\Big(\frac{r}{r_0}\Big)\Big)^4\Big) \ . \nonumber\\ 
\end{eqnarray}
Substituting the above expressions in \eqref{Einstein uu1} we find that it 
does not vanish identically. Since it involves the only unknown quantity $c_5$,
we can use this equation to determine it. Though it is possible reproduce the 
exact expression for $c_5$ in closed form, it is rather complicated and 
unimaginative. Thus we skip writing it here.

\end{document}